\documentstyle[12pt,twoside,epsbox]{article}
{\catcode `\@=11 \global\let\AddToReset=\@addtoreset}
\AddToReset{equation}{section}
\renewcommand{\theequation}{\thesection.\arabic{equation}}  

\def\greaterthansquiggle{\raise.3ex\hbox{$>$\kern-
.75em\lower1ex\hbox{$\sim$}}}
\def\lessthansquiggle{\raise.3ex\hbox{$<$\kern-
.75em\lower1ex\hbox{$\sim$}}}
\newcommand{\beq}{\begin{equation}}
\newcommand{\eeq}{\end{equation}}
\newcommand{\beqa}{\begin{eqnarray}}
\newcommand{\eeqa}{\end{eqnarray}}
\newcommand{\beqan}{\begin{eqnarray*}}
\newcommand{\eeqan}{\end{eqnarray*}}
\newcommand{\ba}{\begin{array}}
\newcommand{\ea}{\end{array}}
\newcommand{\no}{\nonumber}

\newcommand{\Un}{\underline}

\newcommand{\ra}{\rightarrow}

\newcommand{\ve}{\varepsilon}
\newcommand{\vp}{\varphi}

\newcommand{\dg}{\dagger}
\newcommand{\wt}{\widetilde}
\newcommand{\wh}{\widehat}

\newcommand{\A}{{\cal A}}

\newcommand{\E}{{\cal E}}
\newcommand{\F}{{\cal F}}
\newcommand{\G}{{\cal G}}

\newcommand{\M}{{\cal M}}
\newcommand{\N}{{\cal N}}

\newcommand{\R}{{\cal R}}

\newcommand{\dfrac}{\displaystyle \frac}

\def\nz{\ifmmode {I\hskip -3pt N} \else {\hbox {$I\hskip -3pt N$}}\fi}
\def\zz{\ifmmode {Z\hskip -4.8pt Z} \else
       {\hbox {$Z\hskip -4.8pt Z$}}\fi}
\def\qz{\ifmmode {Q\hskip -5.0pt\vrule height6.0pt depth 0pt
       \hskip 6pt} \else {\hbox
       {$Q\hskip -5.0pt\vrule height6.0pt depth 0pt\hskip 6pt$}}\fi}
\def\rz{\ifmmode {I\hskip -3pt R} \else {\hbox {$I\hskip -3pt R$}}\fi}
\def\cz{\ifmmode {C\hskip -4.8pt\vrule height5.8pt\hskip 6.3pt} \else
       {\hbox {$C\hskip -4.8pt\vrule height5.8pt\hskip 6.3pt$}}\fi}

\def\au{{\setbox0=\hbox{\lower1.36775ex%
\hbox{''}\kern-.05em}\dp0=.36775ex\hskip0pt\box0}}
\def\ao{{}\kern-.10em\hbox{``}}

\normalsize
\begin{document}
\bibliographystyle{plain}

\begin{titlepage}
\begin{flushright}
UWThPh-1996-40
\end{flushright}
\vspace{2cm}
\begin{center}
{\Large \bf Nonperturbative Stochastic Quantization of the Helix Model} 
\\[40pt]
Helmuth H{\"u}ffel* and Gerald Kelnhofer** \\
Institut f{\"u}r Theoretische Physik \\
Universit\"at Wien \\
Boltzmanngasse 5, A-1090 Vienna, Austria
\vfill

{\bf Abstract}
\end{center}

The helix model describes the minimal coupling of an abelian 
gauge field with three bosonic matter fields in $0+1$ dimensions; it 
is a model without a global Gribov obstruction. We perform the 
stochastic quantization in configuration space and prove 
nonperturbatively equivalence with the path integral formalism. 
Major points of our approach are the geometrical understanding of 
separations into gauge independent and gauge dependent degrees 
of freedom as well as a generalization of the stochastic gauge 
fixing procedure which allows to extract the equilibrium 
Fokker-Planck probability distribution of the model.

\vfill
\begin{enumerate}
\item[*)] email: helmuth.hueffel@univie.ac.at
\item[**)] supported by ``Fonds zur F\"orderung der wissenschaftlichen 
Forschung in \"Oster\-reich",  project P10509-NAW
\end{enumerate}
\end{titlepage}

\section{Introduction}

An alternative scheme for quantization based on stochastic 
averages has been presented several years ago by Parisi and Wu 
\cite{Parisi+Wu}, see refs. \cite{Damgaard+Huffel,Namiki} for 
comprehensive reviews and referencing. The main idea of 
``stochastic quantization'' in configuration space - in this paper we 
will exclusively discuss the configuration space quantization - is to 
view Euclidean quantum (field) theory as the equilibrium limit of 
a statistical system coupled to a thermal reservoir. This system 
evolves in a new additional time direction which is called 
stochastic time until it reaches the equilibrium limit for infinite 
stochastic time. The coupling to the heat reservoir is simulated by 
means of a stochastic noise field - or rather by the mathematically 
well defined Wiener process - which forces the original Euclidean 
field to wander randomly. In the equilibrium limit the stochastic 
averages become identical to ordinary Euclidean vacuum 
expectation values.

There are two equivalent formulations of stochastic quantization 
due to the general properties of stochastic processes: In one 
formulation all fields have an additional dependence on stochastic 
time. Their stochastic time evolution is determined by a Langevin 
equation which has a drift term constructed from the gradient of 
the classical action of the system. The expectation values of 
observables are obtained by ensemble averages over the Wiener 
measure.

Corresponding to this Langevin equation is an equivalent diffusion 
process which is defined in terms of a Fokker Planck equation for 
the probability distribution characterizing the stochastic evolution 
of the system. Now expectation values of the observables are 
defined by functionally integrating over them with the stochastic 
time dependent Fokker-Planck probability distribution.

In several non-gauge models (see especially 
\cite{Jona-Lasinio+Mitter} for a mathematically rigorous treatment of 
scalar 
field theory) one can verify that in the infinite stochastic time 
limit this Fokker-Planck probability distribution converges to the 
standard Euclidean configuration space path integral density. 

Over the past years the stochastic quantization scheme has grown 
into a useful tool in several areas of quantum field theory and a 
large number of new results and generalized schemes have 
appeared \cite{Damgaard+Huffel+Rosenblum,Namiki+Okano}: In 
specific we would like to recall the applications to gauge fields 
(see below), fermionic fields, the studies of anomalies, 
investigations of supersymmetric models and the regularization 
and renormalization program; last but not the least the numerical 
simulations became an important non perturbative application of 
stochastic quantization; stochastic techniques could be applied to 
Monte-Carlo simulations by means of the so called hybrid 
algorithms.

One might attach great importance especially to the numerical 
applications of the stochastic quantization scheme while supposing 
that its formal field theoretic developments have been completed. 
It is our conviction, however, that the method of stochastic 
quantization can be distinguished by its fundamental concepts and 
therefore seems still relevant for actual research. In this 
connection we refer to the quantization of gauge theories:

In configuration space the most prominent quantization scheme is 
provided by the path integral using the Faddeev-Popov procedure 
\cite{Faddeev+Popov}, which guarantees the unitarity of the 
corresponding Feynman amplitudes. However, due  to the Gribov 
ambiguity \cite{Gribov,Singer}, which is related to the fact that the 
topological nontrivial structure of the space of gauge orbits 
prevents the existence of a global gauge fixing, the path integral 
quantization is incomplete beyond the perturbation theory. On the 
perturbative level this does not lead to difficulties since one stays 
within the local region where the gauge fixing is well defined. 
There have been many attempts to deal with these 
nonperturbative aspects in the path integral 
\cite{Zwanziger90,Shabanov93,Becchi et al,Fujikawa,Friedberg et al.}
 but a profound solution has not 
been found yet.

Much hope has been put forward to gain new insights for a correct 
nonperturbative path integral formulation of gauge theories from 
the stochastic quantization point of view 
\cite{Seiler,Zwanziger86}. The stochastic quantization 
scheme of gauge theories 
in configuration space has been developed apart of the above dual 
structure of a Langevin and Fokker-Planck formulation in two 
related ways: 
\begin{enumerate}
\item[a)] Parisi and Wu \cite{Parisi+Wu} proposed the quantization in 
terms 
of Langevin equations without gauge fixing terms and without 
Faddeev-Popov ghosts. Since the pure action of the gauge model 
remains invariant under gauge transformations, the associated 
drift term of the Langevin equation acts orthogonally to the gauge 
orbits. As a consequence unbounded diffusion along the orbits 
takes place so that potential divergencies are arising for the 
expectation values of general observables. It is argued, however, 
that these contributions are canceling out among themselves in 
expectation values of gauge invariant observables. 

This Parisi-Wu  stochastic quantization method of gauge theories 
preserves the full gauge symmetry and provides a manifestly 
gauge covariant kind of quantization. Due the absence of gauge 
fixing it was claimed to be valid also in the nonperturbative 
regime without the Gribov ambiguity.

\item[b)] In the approach of Zwanziger \cite{Zwanziger81} the drift term 
of 
the 
stochastic process is modified by the addition of a gauge fixing 
force tangent to the gauge orbits. It provides damping for the 
gauge modes' diffusion along the orbits leaving, however, 
unchanged all gauge invariant expectation values. This ``stochastic 
gauge fixing" procedure allows to directly resort to a well defined 
Fokker-Planck formulation so that powerful functional techniques 
can be applied. Again no Faddev-Popov ghosts are needed and no 
Gribov ambiguity was claimed to exist neither.
\end{enumerate}

A fundamental question arises how stochastic quantization - if at 
all - compares with the conventional quantization schemes in the 
case of gauge theories. A great amount of work shows agreement 
at the perturbative level up to some order in the coupling 
constants, e.g. for Yang-Mills theory up to two loops 
\cite{Marculescu+Okano+Schulke+Zheng}. The question of equivalence, 
however, becomes subtle at the perturbative level to any order 
and turns into a hot subject at the nonperturbative level. 

Unfortunately - due to the unbounded diffusion along the gauge 
orbits - the Parisi -Wu scheme for gauge theories lacks an 
immediate Fokker-Planck formulation and equilibrium properties 
are hard to extract. This comes from the technical difficulty to 
solve exactly the corresponding Langevin equations; no general 
equivalence proofs of any nontrivial gauge model have been 
established.

On the other hand using Zwanziger's scheme it is necessary to 
\begin{enumerate}
\item[a)] rigorously control the required stabilizing effect along the 
gauge orbits 
\item[b)] prove convergence to equilibrium of the Fokker-Planck 
probability distribution
\item[c)] explicitly find out what this limit is. 
\end{enumerate}

With respect to these requirements according to our opinion again 
no satisfactory nonperturbative treatment of any nontrivial gauge 
model has been given (see however the formal procedures 
\cite{Baulieu+Zwanziger,Ordonez+Rubin+Zwanziger}).

In view of the above difficulties it is not surprising that the 
interest in the stochastic quantization scheme - despite of its 
conceptional beauties - has gone down remarkably over the last 
years .
 
It seems first of all helpful to choose a gauge model which is as 
simple as possible, containing, however, relevant physical and 
mathematical structure. 

From a pure mathematical viewpoint, a gauge model coupled to 
matter fields can be formulated in terms of fiber bundle theory. 
The fields form an infinite dimensional configuration space on 
which an action of the symmetry group of the model is given. 
Reducing to the true degrees of freedom corresponds to projecting 
the theory down to the true configuration space, which is obtained 
by taking the quotient of the configuration space by the gauge 
group. Depending on the kind of action of the gauge group this 
quotient may have a smooth or an orbifold structure. 

The problem of the orbifold structure of the physical phase space
has been addressed by 
several authors (see e.g. \cite{Emmrich,Tuynman,Gotay}).
 We remind, however, 
 that in this paper we are restricting ourselves 
to just a configuration space discussion. In this respect we would like 
to quote  relations with 
inequivalent quantizations in  two dimensional Yang-Mills theory 
on a cylinder by \cite{Hetrick,Chandar+Ercolessi} and the analysis of
the singularity structure of the Yang-Mills configuration 
space by \cite{Fuchs et al.}. 

In order to gain new insights into the stochastic quantization 
scheme of gauge theories it seems important to analyze how these 
geometrical concepts are encoded in the corresponding Langevin 
or Fokker-Planck equation, respectively. More precisely, we want 
to study the evolution properties of the stochastic process 
separately along the true configuration space as well as along the 
gauge group. In order to avoid orbifold singularities and Gribov 
ambiguities it is appropriate to study such a gauge model in which 
\begin{enumerate}
\item[i)] the gauge group acts freely, that is without fixed points. 

\item[ii)] the separation into gauge independent and gauge dependent 
	degrees of freedom is globally possible.
\end{enumerate}
It is well known (see e.g. \cite{ccc}) that under these circumstances 
the true configuration space is a smooth manifold. Technically, this 
separation corresponds to the choice of a particular trivialization 
of the principal fiber bundle on which the stochastic process is 
formulated.

For these reasons we decided to choose the so called Helix model, which 
describes the minimal coupling of an abelian gauge field with 
three bosonic matter fields in $0+1$ dimensions. This model was 
originally proposed by deWit \cite{deWit}  and was investigated 
intensively within the Hamiltonian framework by Kuchar 
\cite{Kuchar}. Recently the helix model came to new life again 
\cite{Friedberg et al.} in the course of studies on problems with gauge 
fixing and the Gribov ambiguity; it was further analyzed within 
the BRST approach by \cite{Fujikawa}. 

To clarify we remark that although globally the helix model is free 
from topological gauge fixing obstructions one may nevertheless 
find specific gauge choices which suffer from Gribov ambiguities. 
The helix model therefore serves as an ideal testing ground for 
treatments of ``bad" gauge choices and their associated Gribov 
problems. 

In the present paper (see also \cite{StMargh,Physlett})
we perform the stochastic quantization 
procedure by exploiting the globally unproblematic structure of 
the helix model. We will come back to the discussion of the above 
- artificial - complications of the model in a separate publication; 
there we will be analyzing gauge choices where one is forced to 
work on coordinate patches \cite{Huffel+Kelnhofer}.

We present in this paper a refined stochastic gauge fixing 
procedure and prove nonperturbatively the equivalence of the 
stochastically quantized helix model to the known path integral 
formulation. We propose in fact a modified stochastic process 
where not only the drift term is changed but also the Wiener 
process itself. These modifications are done in such a way that a 
proper damping of the gauge modes along the orbits is possible, 
that all gauge invariant expectation values remain unchanged and 
that the equilibrium Fokker-Planck distribution can be 
determined; it is seen to agree with the known path integral 
density of the helix model \cite{Friedberg et al.}. 

Although our generalized scheme of stochastic gauge fixing is 
worked out here for the helix model we will give arguments for 
its general applicability. 

We start in section 2) with a review on basic facts of Ito's 
stochastic  calculus and recall in section 3) essentials of the Parisi-Wu 
stochastic quantization scheme of non-gauge models. 

A discussion on the geometric structure of the helix model in 
configuration space is given in section 4).

In section 5) we present an overview of the stochastic 
quantization of gauge models and derive our generalization of the 
stochastic gauge fixing procedure. 

The damping of the gauge degrees of freedom is discussed in 
section 6).

In section 7) we derive the equilibrium Fokker-Planck 
distribution of the helix model in a nonperturbative manner and 
show that is agreeing with the known path integral density of this 
model. 

Section 8) deals with geometrical aspects of the generalized stochastic 
gauge fixing scheme.

We present an outlook in section 9) and  add a short appendix for technical details.

\section{A short review of the Ito Stochastic Calculus}

In this section we review basic facts of the Ito stochastic calculus 
(see e.g. \cite{Arnold}, \cite{Gardiner}, \cite{Ikeda+Watanabe}) assuming 
that 
the reader is already familiar with the concepts of stochastic 
processes. 

A set of Langevin equations for m scalar fields $\Phi^i(t,s)$, 
$i=1,..,m$,  
which depend on a real coordinate $t$ (which in the following 
sections will be interpreted as Euclidean time) and the stochastic 
time coordinate $s$
\beq
d\Phi^i(t,s) = a^i(t,s)ds + \int_{\bf R} dt' b^i{}_k(t,t';s) dW^k(t',s)
\eeq
is an abreviation for the associated set of stochastic integral 
equations
\beq
\Phi^i(t,s) = \int_{s_0}^s a^i(t,s)d\sigma + \int_{\bf R} dt' 
\int_{s_0}^s b^i{}_k(t,t';\sigma) dW^k(t',\sigma).
\eeq
The $a^i$ and the $b^i{}_k$ are given functions of the fields $\Phi^i$, 
which 
are defined at just the same stochastic time $s$. As a consequence 
the $a^i$ and the $b^i{}_k$ are local in the stochastic time $s$ but are 
allowed to generally be nonlocal in the other variable $t$. Later in 
our paper we will in fact discuss cases where the $a^i$ and the $b^i{}_k$ 
imply differentiations and integrations of the $\Phi^i$ with respect to 
$t$. 

$dW^i(t,s)$ denote the increments of an $m$-dimensional Wiener 
process $W^i(t,s)$, which have the correlations
\beq
\langle\langle dW^i(t,s) dW^k(t',s)\rangle\rangle = 2 \delta^{ik} 
\delta(t-t')ds.
\eeq

The second integral in (2.2) is a stochastic integral; it is defined in 
the mean square sense as the limit of an infinite Riemann-Stieltjes 
sum, which, however, carries a dependence on the choice of the 
positions of the intermediate points. We adopt Ito's choice to 
define the integral according to the product rule
\beq
b^i{}_k(t,t';s)dW^k(t',s) =
\lim_{\Delta s \ra 0} b^i{}_k(t,t';s) [W^k(t,s+\Delta s) - W^k(t,s)]
\eeq
and call the corresponding stochastic differential equation (2.1) an 
Ito-Langevin equation.

Under certain conditions (Lippschitz condition, growth condition, 
specific class of initial conditions) it can be proven that the 
solution of the Ito-Langevin equation (2.1) is a unique 
nonanticipating function. A function $g(s)$ is called non-anticipating 
if for $s < s'$ it is statistically independent of $W(s)-W(s')$. 

Expectation values of arbitrary functions of the solutions $\Phi^i(t,s)$ 
are obtained by ensemble averages over the Wiener measure, 
using the Gaussian behaviour of the Wiener process and applying 
repeatedly the correlations (2.3).

One consequence of Ito's choice is the mean value formula
\beq
\langle\langle \int_{s_0}^s f(\Phi(t,\sigma)) dW^i(t,\sigma)\rangle\rangle
= 0
\eeq
which follows from the non-anticipating nature of the solutions of 
the Ito-Langevin equation; $f$ is an arbitrary (non-anticipating) 
function of $\Phi^i$. 

Next we recall the so called Ito formula, which states that the 
differential of an arbitrary function(al) $f$ of $\Phi^i$ is given by
\beqa
df[\Phi] &=& \int_{\bf R} dt_1 \left\{ \left[ 
\frac{\delta f}{\delta\Phi^i(t_1)} a^i (t_1,s) 
+ \int_{\bf R} dt_2 dt_3 \frac{\delta^2 f}{\delta \Phi^i(t_1)
\delta \Phi^k(t_2)} b^i{}_j(t_1,t_3;s) \delta^{j\ell} 
b^k{}_\ell(t_2,t_3;s)  \right] \right. ds \no \\
&& \left.\left.\mbox{} + \int_{\bf R} dt_2 \frac{\delta f}{\delta 
\Phi^i(t_2)}
b^i{}_k(t_1,t_2;s)dW^k(t_2,s)\right\}\right|_{\Phi(\cdot)=\Phi(\cdot,s)}.
\eeqa
Here it is assumed that $f$ is twice partially continuosly 
differentiable with respect to $\Phi^i$. 

Equivalently to (2.6) we can formally define multiplication rules 
for the differentials $dW^i$ 
\beqa
dW^i(t,s) dW^k(t',s) &=& 2 \delta^{ik} \delta(t-t') ds \no \\
dW^i(t,s)^{2+N} &=& 0 \qquad N = 1,2,\ldots \\
dW^i(t,s)ds &=& 0, \no
\eeqa
which -- more precisely -- are given in 
terms of associated integrals over arbitrary non-anticipating 
functions. From (2.6) or (2.7) amounts, unfortunately, the cumbersome 
conclusion that in calculating expressions involving infinitesimals 
$ds$ of the stochastic time one must keep all expansions up to the 
second order in the $dW^i$. 

What usually is meant by Ito's stochastic calculus can be 
summarized by equations (2.4) -- (2.7).

It is of fundamental significance that with the help of Ito's as well 
as the mean value formula one can derive from the Ito-Langevin 
equation (2.1) the Fokker-Planck equation for a (conditional) 
probability density $\rho$
\beq
\frac{\partial \rho[\Phi,s]}{\partial s} = L[\Phi] \rho[\Phi,s].
\eeq
Here the so called Fokker-Planck operator $L$ is defined in terms of 
the $a^i$ and the $b^i{}_k$ as
\beq
L[\Phi] = \int_{\bf R} dt_1 \frac{\delta}{\delta\Phi^i(t_1)}[-a^i(t_1) +
\int_{{\bf R}^2} dt_2 dt_3 \frac{\delta}{\delta \Phi^k(t_2)}
(b^i{}_j(t_1,t_3) \delta^{j\ell} b^k{}_\ell(t_2,t_3))].
\eeq
It should be stressed that in a Fokker-Planck formulation the 
fields $\Phi^i$ - and consequently also the $a^i$ and the $b^i{}_k$ - are 
stochastic time independent; all the stochastic time dependence is 
coming from the probability density $\rho$ itself. We therefore 
omit to denote a $s$-dependence of $\Phi^i$, as well as of $a^i$ and 
$b^i{}_k$.

Due to the fundamental relationship between the Ito-Langevin 
(2.1) and the Fokker-Planck (2.8, 2.9) scheme  the expectation 
values of arbitrary (equal stochastic time) functions $f$ of the 
solutions $\Phi^i(t,s)$ of the Ito-Langevin equation are coinciding 
with expectation values defined correspondingly in terms of the 
Fokker-Planck density $\rho$
\beq
\langle\langle f(\Phi(\cdot,s))\rangle\rangle =
\langle f(\Phi(\cdot))\rangle(s) =
\int D\Phi \; f(\Phi(\cdot)) \rho[\Phi,s].
\eeq
We remark again that in (2.10) there are involved stochastic time 
dependent as well as stochastic time independent fields, 
respectively. 

The Fokker-Planck operator $L$ describes the stochastic time 
evolution of the Fokker-Planck probability density; with its help 
we formally write 
\beq
\rho[\Phi,s] = e^{sL[\Phi]} \rho[\Phi,s=0].
\eeq
In case that we are concerned with the stochastic time evolution 
of expectation values, it is advantageous, however, to introduce in 
addition to $L$ also its adjoint $L^\dg$
\beq
\int D\Phi \; f(\Phi(\cdot)) \rho[\Phi,s] = 
\int D\Phi (e^{sL[\Phi]} \rho[\Phi,s=0]) f(\Phi(\cdot)) =
\int D\Phi \; \rho[\Phi,s=0] e^{sL^\dg[\Phi]} f(\Phi(\cdot)).
\eeq
It explicitly reads in terms of  the $a^i$ and the $b^i{}_k$
\beq
L^\dg[\Phi] = \int_{\bf R} dt_1 \left[ a^i(t_1) +
\int_{{\bf R}^2} dt_2 dt_3 b^i{}_j (t_1,t_3) \delta^{j\ell} b^k{}_\ell
(t_2,t_3) \frac{\delta}{\delta\Phi^k(t_2)} \right]
\frac{\delta}{\delta \Phi^i(t_1)}.
\eeq
At the end of this section we remark that we discussed Ito's 
stochastic calculus for fields $\Phi^i$ depending on just the 
coordinate(s) $t$ (and $s$), having in mind applications to the helix 
model. It would otherwise have been straightforward to 
generalize this section to a higher dimensional case as well.

\section{A short review of the Parisi-Wu stochastic quantization scheme}

In this section we review some of the important features of the 
stochastic quantization scheme when performing the quantization 
of non-gauge models. We include this section for the benfit of the 
reader to get a self-contained presentation of our paper in order to 
be able to follow closely our subsequent discussions of the 
stochastic quantization procedure in the case of gauge theories.

We consider a model of $m$ (self-interacting) scalar fields $\Phi^i$ with 
the positive Euclidean action denoted by $S$. We assume that the 
model has a finite normalization integral $\int D\Phi \; e^{-S}$.

The basic idea of the stochastic quantization scheme is to interpret 
the Euclidean path integral measure $e^{-S}/ \int D\Phi \; e^{-S}$ as 
the equilibrium limit of a Fokker-Planck probability distribution 
of a specific stochastic process. 

This stochastic process is defined by an Ito-Langevin equation 
(2.1) of the following form: The drift term $a^i$ is taken to be the 
variation of the action $S$ with respect to the fields $\Phi^i$, whereas 
the matrix $b^i{}_k$ is chosen most simply as an $m$-dimensional unit 
matrix.
\beq
d\Phi^i(t,s) = - \left. \delta^{ij} \frac{\delta S}{\delta \Phi^j(t)}
\right|_{\Phi(t) = \Phi(t,s)} ds + dW^i(t,s).
\eeq

It then follows from (2.9) that the associated Fokker-Planck 
operator is just
\beq
L[\Phi] = \int_{\bf R} dt \frac{\delta}{\delta \Phi^i(t)} \delta^{ij} \left[
\frac{\delta S}{\delta \Phi^j(t)} + \frac{\delta}{\delta \Phi^j(t)}
\right].
\eeq
We nonperturbatively deduce that the equilibrium limit of the 
Fokker-Planck distribution is formally given by 
$e^{-S}/\int D \Phi  \; e^{-S}$. 
In this paper we do not discuss the complications due to 
the issue of renormalization, see, however \cite{Jona-Lasinio+Mitter} or 
\cite{Marculescu+Okano+Schulke91}.

As a consequence the equal stochastic time expectation values of 
arbitrary observables (calculated either within the Langevin or 
Fokker-Planck approach, (2.10)) relax in the infinite stochastic time 
limit 
to the corresponding quantum Green functions
\beq
\lim_{s \ra \infty}
\langle\langle f(\Phi(\cdot,s))\rangle\rangle = \lim_{s \ra \infty}
\langle f(\Phi(\cdot))\rangle(s) = \langle f(\Phi(\cdot))\rangle.
\eeq

For later application we study now a  bijective change of variables of
 the original fields 
 $\Phi^i(t)$ to new fields $\Psi^\mu(t)$ \cite{Graham,Caracciolo 
et 
al.,Batrouni et al.}. With respect to this variable 
change we introduce vielbeins $E$ and their inverses $e$
\beq
E^\mu{}_i(t,t') = \frac{\delta \Psi^\mu(t)}{\delta \Phi^i(t')}, \qquad
e^i{}_\mu(t,t') = \frac{\delta \Phi^i(t)}{\delta \Psi^\mu(t')}, 
\eeq
as well as the induced inverse metric $G^{\mu\nu}$ and 
the determinant $G$
\beqa
G^{\mu\nu}(t_1,t_2) &=& \int_{\bf R} dt_3 \; E^\mu{}_i(t_1,t_3)
\delta^{ij} E^\nu{}_j(t_2,t_3) \no \\
G &=& \det G_{\mu\nu}(t_1,t_2).
\eeqa
The vielbeins,  the inverse metric $G^{\mu\nu}$ and the 
determinant $G$ are generally field dependent and are used either 
in the Fokker-Planck or in the Langevin equations; in the latter 
case we have to explicitly insert the additional $s$-dependence of 
the fields. 

With the Ito formula (2.6) the transformed Langevin equations for 
the new variables $\Psi^\mu$ read
\beq
d\Psi^\mu(t,s) = \int_{\bf R} dt' \left\{ \left. E^\mu{}_i(t,t')
\left[ -\frac{\delta S}{\delta \Phi^i(t')}ds + dW^i(t',s)\right]
+ \frac{\delta E^\mu{}_i(t,t')}{\delta \Phi^i(t')} ds 
\right\} \right|_{\Phi(\cdot)=\Phi(\cdot,s)}
\eeq
which we can recast into
\beqa
d\Psi^\mu(t,s) &=& 
 \int_{\bf R} dt' \left\{ \left[ - G^{\mu\nu}(t,t')
\frac{\delta S}{\delta \Psi^\nu(t')} 
+ \frac{1}{\sqrt{G}} \; \frac{\delta}{\delta \Psi^\nu(t')}
(G^{\mu\nu}(t,t') \sqrt{G}) \right] \right. ds \no \\
&& \left. \left. \mbox{} + E^\mu{}_i(t,t')dW^i(t',s) 
\right\} \right|_{\Psi(\cdot)=\Psi(\cdot,s)}. 
\eeqa
In the last equation we made use of the fact that the covariant 
divergence of the vielbein $E$ is vanishing (we considered the 
transformations from flat  field variables to curved ones) so that 
\beq
\int_{\bf R} dt \left[ \frac{\delta E^\mu{}_i(t,t')}{\delta \Psi^\mu(t)}
+ \frac{1}{\sqrt{G}} \; E^\mu{}_i(t,t')
\frac{\delta \sqrt{G}}{\delta \Psi^\mu(t)} \right] = 0
\eeq
and
\beq
\int_{\bf R} dt' \frac{\delta E^\mu{}_i(t,t')}{\delta \Phi^i(t')}
= \int_{\bf R} dt' 
\frac{1}{\sqrt{G}} \; \frac{\delta}{\delta \Psi^\nu(t')}
(G^{\mu\nu}(t,t') \sqrt{G}).
\eeq
We interpret (3.7) as a stochastic process taking place in the 
manifold parametrized by the metric $G_{\mu\nu}$, see e.g. \cite{Graham}; 
we observe the 
presence of the so called ``Ito term'', which appears as a simple 
consequence of the Ito formula.

We are interested in the transformed form of the Fokker-Planck 
equation, too. We have just to use (2.9) in the case of the new 
variables $\Psi^\mu$, starting from the transformed Langevin 
equation (3.7). We immediately obtain the Fokker-Planck operator 
in the new variables as
\beq
L[\Psi] = \int_{{\bf R}^2} dt dt' \frac{\delta}{\delta \Psi^\mu(t)}\;
G^{\mu\nu}(t,t') \left[ \frac{\delta S}{\delta \Psi^\nu(t')} -
\frac{1}{\sqrt{G}} \; \frac{\delta \sqrt{G}}{\delta \Psi^\nu(t')}
+ \frac{\delta}{\delta \Psi^\nu(t')}\right].
\eeq
The formal equilibrium distribution now reads
\beq
\rho[\Psi]_{\rm equil.} = \frac{\sqrt{G} \; e^{-S}}
{\int D \Psi \; \sqrt{G} \; e^{-S}}
\eeq
and we observe the appearance of the familiar Jacobian factor 
$\sqrt{G}$ corresponding to the variable transformation $\Phi^i$ to 
$\Psi^\mu$.

Finally we comment on an equivalence relation of stochastic 
processes in the infinite stochastic time limit, which is usually 
called fluctuation-dissipation theorem, or sometimes kernel 
independence of the stochastic process (see  e.g. \cite{Gardiner} and 
\cite{Okano et al. 91}). More 
precisely we consider Ito-Langevin equations of the form  
\beqa
d\Psi^\mu(t,s) &=& 
 \int_{\bf R} dt' \left\{ \left[ - G^{\mu\nu}(t,t')
\frac{\delta S}{\delta \Psi^\nu(t')} 
+  \frac{\delta}{\delta \Psi^\nu(t')}
G^{\mu\nu}(t,t')  \right] \right. ds \no \\
&& \left. \left. \mbox{} + E^\mu{}_i(t,t')dW^i(t',s) 
\right\} \right|_{\Psi(\cdot)=\Psi(\cdot,s)}. 
\eeqa
which are distinguished from (3.7) with respect to a different 
form of the Ito term. We remark that $G^{\mu\nu}$ is assumed  to be  
positive as above with the determinant $G$ being arbitrary. It 
follows from (3.12) that $G^{\mu\nu}$ factorizes in the corresponding 
Fokker-Planck operator
\beq
L[\Psi] = \int_{{\bf R}^2} dt dt' \frac{\delta}{\delta \Psi^\mu(t)}\;
G^{\mu\nu}(t,t') \left[ \frac{\delta S}{\delta \Psi^\nu(t')} 
+ \frac{\delta}{\delta \Psi^\nu(t')}\right]
\eeq
so that the equilibrium distribution is given independently of the 
explicit form of $G^{\mu\nu}$ by
\beq
\rho[\Psi]_{\rm equil.} = \frac{e^{-S}}{\int D\Psi \; e^{-S}}.
\eeq

\section{The geometrical structure of the helix model}

The helix model is defined by the Lagrange density
\beq
L(t) = \frac{1}{2} [(\dot \vp^1(t) - A(t) \vp^2(t))^2 +
(\dot \vp^2(t) + A(t) \vp^1(t))^2 +
(\dot \vp^3(t) - A(t))^2] -
\frac{1}{2} [(\vp^1(t))^2 + (\vp^2(t))^2]
\eeq
where the dot denotes time derivation and the fields
$$
\vec \vp(t) = (\Un{\vp}(t),\vp^3(t)) = (\vp^1(t),\vp^2(t),\vp^3(t))
$$
and $A(t)$ are regarded as elements of the function spaces 
$\E = C^\infty({\bf R},{\bf R}^3)$ and $\A = C^\infty({\bf R},{\bf R})$,
respectively.

Let $\G = C^\infty({\bf R},{\bf R})$ denote the abelian group of local
gauge transformations where $(g_1 \cdot g_2)(t) = g_1(t) + g_2(t)$ gives 
the
group structure. Consider the following transformation $\R$ of $\G$
on the configuration space $\E \times \A$
\beq
\R(\vec \vp,A,g)(t) = (R(g(t)) \Un{\vp}(t), \vp^3(t) - g(t),A(t) - 
\dot g(t)),
\eeq
where $g \in \G$ and
$$
R(g(t)) = \left( \ba{rr} 
\cos g(t) & - \sin g(t) \\ \sin g(t) & \cos g(t) \ea \right).
$$
The Lagrange density (4.1) is easily verified to be invariant under these
transformations and therefore it descends to a function on the quotient
space $\E \times_\G \; \A := (\E \times \A)/\G$ which is obtained by
factoring out the gauge degrees of freedom. The generator of infinitesimal
gauge transformations is given by the vector field $Z_\xi(\Phi)$ with 
components $Z^i(t',t)$
\beqa
Z_\xi(\Phi) &=& \int_{\bf R} dt \left[ -\xi(t) \vp^2(t) 
\frac{\delta}{\delta \vp^1(t)} + \xi(t) \vp^1(t)\frac{\delta}
{\delta \vp^2(t)} - \xi(t) \frac{\delta}{\delta \vp^3(t)} -
\dot \xi(t) \frac{\delta}{\delta A(t)} \right] \no\\
&=&\int _{{\bf R}^2} dt dt' \xi(t') Z^i(t',t) \frac{\delta}{\delta 
\Phi^i(t)} 
\eeqa
on $\E \times \A$, where $\xi \in \mbox{Lie }\G = C^\infty({\bf R},{\bf 
R})$
is an arbitrary element of the Lie algebra of the gauge group and where 
we defined
$$
\Phi(t)=(\vec \vp(t),A(t)) = (\vp^1(t),\vp^2(t),\vp^3(t),A(t)).
$$ 
The
transformation properties (4.2) suggest to view the fields $A(t)$ and
$\vp(t)$ as gauge field and matter fields, respectively, so that the
helix model can be interpreted as an example for a $0+1$ dimensional
gauge theory on the real line {\bf R} coupled with three matter fields.
From a more mathematical point of view, the space $\A$ can be identified
with the affine space of connections on the trivial principal fiber
bundle over the base {\bf R} with the additive group of real numbers as
corresponding structure group and the matter fields can then be realized
as sections of a particular vector bundle with fiber ${\bf R}^3$
associated to that given principal bundle.

The true configuration space $\E \times_\G \; \A$ is a smooth manifold
which follows from the fact that the gauge group $\G$ acts freely on
$\E \times \A$, which means that the equation $\R(\vec \vp,A,g) = 
(\vec \vp,A)$ for arbitrary fields $(\vp,A) \in \E \times \A$ only admits
the trivial solution $g = 0$. Moreover the projection
$\pi : \E  \times \A \ra \E \times_\G \; \A$, with
\beq
\pi(\vec \vp,A) =
[\vec \vp,A],
\eeq
where $[\vec \vp,A]$ denotes the equivalence class of
$(\vec \vp,A)$ with respect to the action $\R$, provides the structure
of a principal fiber bundle with total space $\E \times \A$, base
manifold $\E \times_\G \; \A$ and structure group $\G$.

In this paper we aim at a nonperturbative application of the stochastic
quantization scheme to the helix model. In order to proceed it will
turn out to be of great importance to rewrite the stochastic process in 
terms of gauge invariant and gauge dependent fields. However, from a
mathematical viewpoint, such a rewriting really can be performed if
and only if the principal fiber bundle $\pi : \E \times \A \ra \E 
\times_\G \; \A$ is trivializable. In this case every field configuration
$(\vec \vp,A)$ can be represented in terms of coordinates parametrizing
$\E \times_\G \; \A$ and $\G$, respectively.

In order to prove the triviality of the bundle we necessarily have to
find a global section, that means a map $\sigma : \E \times_\G \; \A
\ra \E \times \A$ satisfying $\pi \circ \sigma = id$. We will study two
 specific examples for such global sections. 

First we can easily verify that $\sigma_1 : \E \times_\G \; \A \ra \E \times \A$
given by
\beq
\sigma_1([\vec \vp,A]) = (R(\vp^3)\Un{\vp},0,A - \dot \vp^3)
\eeq
is indeed a global section, corresponding in physical terms to the axial gauge
$\vp^3 = 0$.

However we shall display
another global section of the bundle which implies the temporal
gauge $A = 0$. For this case let us take an arbitrary but fixed function $h \in C^\infty
({\bf R},{\bf R})$ with the property that $\int_{\bf R} dt  h(t) = 1$ and
$\int_{\bf R} dt  h(t)^2 =1$. The map $\sigma_2 : \E \times_\G \;
\A \ra \E \times \A$ given by
\beq
\sigma_2([\vec \vp,A]) = (R(F(\vec \vp,A))\Un{\vp},\vp^3 - F(\vec \vp,A),0)
\eeq
where $F: \E \times \A \ra \G$ is defined by
\beq
F(\vec \vp,A)(t) := \int_{\bf R} dt' h(t')
\left[ \int_{t'}^t dt'' A(t'') + \vp^3(t')\right]
\eeq
yields a global section of $\pi : \E \times \A \ra \E \times_\G \; \A$.
To see this, we first note that $\sigma_2$ is well defined which follows
from $F(R(\vec \vp,A,g)) = F(\vec \vp,A) - g$. To prove finally that
$\sigma_2$ is a section we have to use the fact that
$\dot F(\vec \vp,A) = A$.

For a given section $\sigma_i$ we have a diffeomorphism $\chi_i : (\E \times_\G \; \A)
\times \G \ra \E \times \A$ given by
\beq
\chi_i([\vec \vp,A],g) = \R(\sigma_i([\vec \vp,A]),g)
\eeq
where $i = 1,2$. Let $\omega_{(\sigma_i)} : \E \times \A \ra \G$ be the function which is
uniquely determined by the equation
\beq
(\vec \vp,A) = \R(\sigma_i(\pi(\vec \vp,A)),\omega_{(\sigma_i)}(\vec \vp,A)).
\eeq
Explicitely we find
\beq
\omega_{(\sigma_i)}(\vec \vp,A) = \sigma^3_i([\vec \vp,A])- \vp^3,
\eeq
where $\sigma^3_i([\vec \vp,A])$ denotes the third component of the i-th section $\sigma_i$.
For the two choices (4.5) and (4.6) for the  sections $\sigma_i$ we have
\beq
\omega_{(\sigma_1)}(\vec \vp,A) = -\vp^3
\eeq
and
\beq
\omega_{(\sigma_2)}(\vec \vp,A) = - F(\vec \vp,A),
\eeq
respectively.
We remark that under a gauge transformation $\omega_{(\sigma_i)}$ transforms simply like
\beq
\omega_{(\sigma_i)}(\R(\vec \vp,A,g)) = \omega_{(\sigma_i)}(\vec \vp,A) + g
\eeq
and for later use  also mention the important property
\beq
\omega_{(\sigma_i)}(\sigma_i([\vec \vp,A])) = 0.
\eeq

Now it can  be verified straightforwardly that the inverse mapping
$\chi_i^{-1} : \E \times \A \ra (\E \times_\G \; \A) \times \G$
is given by
\beq
\chi_{(i)}^{-1}(\vec \vp,A) = ([\vec \vp,A],\omega_{(\sigma_i)}
(\vec \vp,A))
\eeq
and it is therefore globally possible to rewrite every field configuration
in the appropriate manner. In physical terms the triviality of the
principal fiber bundle $\pi : \E \times \A \ra \E \times_\G \; \A$
means that the helix model does not suffer from a topological Gribov
problem. 

However, for further use it is inconvenient to rewrite all fields in
terms of equivalence classes on $\E \times_\G \; \A$. Therefore it seems
to be more appropriate to introduce coordinates which parametrize a
space diffeomorphic to the orbit space $\E \times_\G \; \A$ which
also may be embedded into the original configuration space
$\E \times \A$. A natural choice is provided by the image of the global
section $\sigma$.

Let $\Sigma_{(\sigma_i)} = \mbox{im } \sigma_i$ denote the gauge fixing 
surface
in $\E \times \A$ for a given section $\sigma_i$ then we can consider
the isomorphic principal fiber bundle
$\pi_{(\sigma_i)} : \E \times \A \ra \Sigma_{(\sigma_i)}$ with projection
$\pi_{(\sigma_i)} = \sigma_i \circ \pi$,
\beq
\pi_{(\sigma_i)}(\vec \vp,A) = \sigma_i([\vec \vp,A]),
\eeq
 total space $\E \times \A$, 
and base manifold $\Sigma_{(\sigma_i)}$, which can be treated as a 
submanifold
in $\E \times \A$ (see Fig.~1 for these geometrical structures).

At this point 
we
want to remark that due to (4.14)  the fields $(\vec \vp,A)$ on the gauge fixing surface  $\Sigma_{(\sigma_i)}$
can equivalently be characterized by the implicit condition
\beq
\omega_{(\sigma_i)}(\vec \vp,A) = 0.
\eeq
For our first choice (4.5) of the global section $\sigma_1$  we simply obtain from (4.11), (4.17) that $\vp^3=0$, whereas for   $\sigma_2$ defined in (4.6) we derive
\beq
A=0 \ \ \ {\rm as \ well \ as} \ \ \  \int_{\bf R} dt h(t) \vp^3(t)=0
\eeq
 for 
$h$ with the properties as given above. 

Corresponding to the  bundle
$\pi_{(\sigma_i)} : \E \times \A \ra \Sigma_{(\sigma_i)}$ a section can be chosen to be the identity mapping so that in analogy to (4.8)  the bundle trivialization
$\chi_{(\sigma_i)} : \Sigma_{(\sigma_i)} \times \G \ra \E \times \A$ is 
given by
\beq
\chi_{(\sigma_i)}(\sigma_i([\vec \vp,A]),g) = \R(\sigma_i([\vec \vp,A]),g).
\eeq

Corresponding to (4.10) and (4.15) the inverse mapping $\chi_{(\sigma_i)}^{-1} : \E \times \A \ra
\Sigma_{(\sigma_i)} \times \G$ reads 
\beq
\chi_{(\sigma_i)}^{-1}(\vec \vp,A) = (\sigma_i([\vec \vp,A]),
\omega_{(\sigma_i)}(\vec \vp,A)).
\eeq
In subsequent chapters we will use exactly this diffeomorphism
$\E \times \A \cong \Sigma_{(\sigma_i)} \times \G$ for studying the
stochastic process associated with the helix model. This result
provides the way to separate the original process in the configuration
space $\E \times \A$ into two subprocesses, one of which is formulated  just in terms of fields $(\vec \vp,A) \in \Sigma_{(\sigma_i)}$ whereas the other process is involving gauge dependent fields as well. Since this 
diffeomorphism
explicitly depends on the chosen section $\sigma_{(i)}$, the
corresponding processes will be related with the actual gauge fixing.

Let us introduce general coordinates $\vec \Psi_{(\sigma_i)}$ on $\Sigma_{(\sigma_i)}$
$$
\vec \Psi_{(\sigma_i)} = (\Un{\Psi}_{(\sigma_i)},\Psi^3_{(\sigma_i)})
= (\Psi^1_{(\sigma_i)}, \Psi^2_{(\sigma_i)}, \Psi^3_{(\sigma_i)})
$$
 and a coordinate $\Psi^4_{(\sigma_i)}$
parametrizing $\G$, which we combine into
$$
\Psi_{(\sigma_i)}=(\vec \Psi_{(\sigma_i)},\Psi^4_{(\sigma_i)}) =  
(\Psi^1_{(\sigma_i)}, \Psi^2_{(\sigma_i)}, \Psi^3_{(\sigma_i)}, 
\Psi^4_{(\sigma_i)}).
$$ 
 From (4.20) we trivially deduce $3$-dimensional coordinates on $\Sigma_{(\sigma_i)}$ by observing that for each of the two choices (4.5), (4.6)
the $4$-vector $\sigma_i([\vec \vp,A])$  has one vanishing component;  $\Psi^4_{(\sigma_i)}$ is taken to be $\omega_{(\sigma_i)}(\vec \vp,A)$. 

 For the
particular section $\sigma_1$ in (4.5), (4.20) yields
\beqa
\Un{\Psi}_{(\sigma_1)} &=& R(\vp^3) \; \Un{\vp} \no \\
\Psi^3_{(\sigma_1)} &=& A - \dot \vp^3 \\
\Psi^4_{(\sigma_1)} &=& \omega_{(\sigma_1)}(\vec \vp,A) = - \vp^3 \no
\eeqa
with inverse 
\beqa
\Un{\vp} &=& R(\Psi^4_{(\sigma_1)}) \; \Un{\Psi}_{(\sigma_1)} \no \\
\vp^3 &=& -\Psi^4_{(\sigma_1)} \\
A &=& \Psi^3_{(\sigma_1)} - \dot \Psi^4_{(\sigma_1)}. \no
\eeqa
For $\sigma_2$ as defined in (4.6) the corresponding coordinate transformation (4.20) reads
\beqa
\Un{\Psi}_{(\sigma_2)} &=& R(F(\vec \vp,A)) \; \Un{\vp} \no \\
\Psi^3_{(\sigma_2)} &=& \vp^3 - F(\vec \vp,A) \\
\Psi^4_{(\sigma_2)} &=& -F(\vec \vp,A) \no
\eeqa
with inverse
\beqa
\Un{\vp} &=& R(\Psi^4_{(\sigma_2)}) \; \Un{\Psi}_{(\sigma_2)} \no \\
\vp^3 &=& \wt \Psi^3_{(\sigma_2)} - \Psi^4_{(\sigma_2)} \\
A &=& -\dot \Psi^4_{(\sigma_2)} \no
\eeqa
where
\beq
\wt \Psi^3_{(\sigma_2)}(t) = \int_{\bf R} dt' \left[\delta(t-t') -h(t) h(t')\right] \Psi^3_{(\sigma_2)}(t').
\eeq
Notice that corresponding to (4.17) and (4.18) the coordinates $\vec \Psi_{(\sigma_2)}$ on 
$\Sigma_{(\sigma_2)}$ fulfill  $\int_{\bf R} dt h(t) \Psi^3_{(\sigma_2)}(t)=0$ which explains  the presence of the projection operator $\delta(t-t') - h(t) h(t')$ acting on $\Psi^3_{(\sigma_2)}$ in (4.24) and (4.25). This operator is not needed in (4.23) as $\Psi^3_{(\sigma_2)}$ defined in terms of $(\vec \vp,A)$ automatically fulfills the required relation .

In the new coordinates, gauge transformations are given purely as
translations, i.e. 
$(\vec \Psi_{(\sigma_i)},\Psi^4_{(\sigma_i)}) \ra
(\vec \Psi_{(\sigma_i)},\Psi^4_{(\sigma_i)} + g)$
where $g \in \G$. Correspondingly, the generator for an infinitesimal
gauge transformation $\xi \in \mbox{Lie }\G$
\beq
Z_\xi(\Psi_{(\sigma_i)}) =
\int_{\bf R} dt \; \xi(t) \frac{\delta}{\delta \Psi^4_{(\sigma_i)}(t)}
\eeq
is just the invariant vector field along the gauge group.

Let us discuss our findings of this chapter in more detail:
We have seen that the principal fiber bundle associated with the helix
model is trivializable and so does not suffer from a Gribov problem. To
clarify this point we recall that in strict mathematical terms (as
analyzed by Singer \cite{Singer} in the case of Yang-Mills theory) one
speaks about a Gribov problem if the projection from the original
configuration space onto the space of gauge orbits gives a 
non-trivializable principal fiber bundle. In this case it is not
possible to find a separation of any field configuration in terms
of gauge invariant and gauge variant fields in a global way. Such a
separation can be done only locally and as a result there may occur
ambiguities in the choice of appropriate coordinates on the local
patches of the orbit space which have nonvoid intersection.

Our main idea was to select a gauge fixing surface in the original
configuration space. This choice was naturally given by taking a
particular section in the corresponding principal fiber bundle.
The gauge fixing surface was then characterized as a set of solution
of a certain function (4.11) on the configuration space depending,
in fact, on the chosen section.

We recall at this place that recently several papers have been
published \cite{Friedberg et al.,Fujikawa} which take the helix 
model as an (explicitly
soluble) example for analyzing the Gribov problem. The apparent
discrepancy to our result can be traced back to different concepts of
definitions:
There exists of course the possibility to define an arbitrary function
on the configuration space, consider the set of points where it vanishes 
and ask if this set is a diffeomorphic copy of the gauge orbit space. It is
precisely this route of analysis which has been followed in the above
mentioned papers. Depending on the actual form of the given function, it
may be possible to find at least a certain subset of solutions which is
diffeomorphic to a subset of the orbit space. In that case the theory
was said to possess a Gribov ambiguity. We prefer to interpret the
kind of Gribov ambiguity which is talked about in this context as
being of an artificial kind.

In a following paper \cite{Huffel+Kelnhofer} 
we shall present a stochastic quantization 
analysis of this artificial Gribov problem associated with sections of
the bundle $\pi : \E \times \A \ra \E \times_\G \; \A$ which can be
defined only locally.

\section{Generalized stochastic gauge fixing}

In this section we review the stochastic quantization procedure of 
gauge (field) theories and present our generalisation of the 
stochastic gauge fixing procedure of Zwanziger \cite{Zwanziger81}.

In order to work out the new features of our modified approach 
we  review at the beginning the Parisi-Wu stochastic quantization 
scheme \cite{Parisi+Wu} for gauge theories and Zwanziger's original 
gauge fixing procedure. This summary serves as a complement to 
our general introductory remarks in sect. 1) and to our short 
review in sect. 3) of the stochastic quantization of non-gauge 
models. 

We collectively denote by $\Phi^k$ the pure gauge, as well as matter 
fields of the given gauge model. For reasons of simplicity we 
choose a notation in correspondance with the helix model; in 
specific we introduce 4 fields depending on the (Euclidean) time 
coordinate $t$, as the helix model is defined in terms of 4 fields in
 $0+1$ dimensions. We 
remark that a generalization to more space time dimension and to 
more general gauge models is immediate. According to the setting 
of stochastic quantization within the Langevin picture the fields 
are defined to depend additionally on the stochastic time 
coordinate $s$ as well; explicitly this amounts to substitute $\Phi^k(t)$ 
by $\Phi^k(s,t)$. 

The crucial point of the Parisi-Wu approach for the gauge theory 
case is to demand that the stochastic time evolution of the fields is 
given again by a Langevin equation of the form
\beq
d\Phi^i(t,s) = - \left. \delta^{ij} \frac{\delta S}{\delta \Phi^j(t)}
\right|_{\Phi(t) = \Phi(t,s)} ds + dW^i
\eeq
where $S$ denotes the original (Euclidean) action of the gauge 
model; it is the unmodified bare action without gauge symmetry 
breaking terms and without accompanying ghost field terms; in 
our case of the helix model it is given by the time integral of the 
Wick rotated form of the Lagrangian (4.1). 

We have alread remarked in the introductiory section that in the case of 
a gauge theory the 
stochastic process defined by (5.1) undergoes undamped diffusion 
and does not approach an equilibrium distribution. Related to this 
fact is that a Fokker-Planck formulation for the $\Phi^k$ is not 
possible because the gauge invariance of the action leads to 
divergencies in the normalization condition of the Fokker-Planck 
density. 

We now discuss Zwanziger's modified formulation \cite{Zwanziger81} of 
the Parisi-Wu scheme: The stochastic gauge fixing procedure 
consists in adding an additional drift force to the Langevin 
equation (5.1) which acts tangentially to the gauge orbits. This 
additional term generally can be expressed by the gauge 
generator and an arbitrary function $\alpha$; in the specific case of 
the helix model the components of the gauge generator are defined 
by (4.3) so that the modified Langevin equation reads as follows
\beq
d\Phi^i(t,s) = - \left. \left[ \delta^{ij}  \frac{\delta S}{\delta \Phi^j(t)} +
\int_{\bf R} dt'Z^i(t,t')\alpha(t')\right]
\right|_{\Phi(\cdot) = \Phi(\cdot,s)} ds + dW^i
\eeq
Let us note two remarkable consequences of the extra term in 
(5.2) 
\begin{enumerate}
\item[a)] The expectation values of gauge invariant observables remain 
unchanged for any choice of the function $\alpha$ (see below for the 
explicit demonstration contained in the discussion of our 
generalized stochastic gauge fixing procedure)

\item[b)] For specific choices of the -- in principle -- arbitrary 
function 
$\alpha$ the gauge modes' diffusion is damped along the gauge 
orbits. As a consequence the Fokker-Planck density can be 
normalized and arbitrary expectation values of the fields are 
relaxing exponentially in the stochastic time s to equilibrium 
values. We remind that this situation is in contrast to the Parisi-
Wu approach, where for expectation values of gauge variant 
observables no equilibrium values could be attained.
\end{enumerate}

We present now our generalization of Zwanziger's stochastic gauge 
fixing procedure by adding a specific drift term 
which  not only has tangential components along 
 the gauge orbits; in addition we modify the 
Wiener process itself. Due to this generalization we are able to 
introduce more than just only one extra function $\alpha$, in fact we 
additionally introduce 4 arbitrary functions $\beta_k$. 
Our generalization is done in such a way that 
expectation values of gauge invariant observables again remain 
untouched. 

Our generalized Langevin equation reads
\beqa
d\Phi^i(t,s) &=& -  \left[ \delta^{ij}  \frac{\delta S}{\delta \Phi^j(t)} +
\int_{\bf R} dt'Z^i(t,t')\alpha(t')\right. \no \\
&& \mbox{} + \left.\left. \int_{{\bf R}^2} dt_1 dt_2
\frac{\delta Z^i(t,t_1)}{\delta \Phi^k(t_2)} \; \zeta^k(t_1,t_2)
\right] \right|_{\Phi(\cdot) = \Phi(\cdot,s)} ds \no \\
&& \mbox{} + \left. \int_{\bf R}dt_2 \left[\delta^i{}_k \delta(t-t_2)
+ \int_{\bf R} dt_1 Z^i(t,t_1) \beta_k(t_1,t_2)\right]
\right|_{\Phi(\cdot) = \Phi(\cdot,s)} dW^k(t_2,s) \no \\
\eeqa
here we introduced $\zeta^k(t_1,t_2)$ as a shorthand notation of
\beq
\zeta^k(t_1,t_2) = 2 \delta^{k\ell} \beta_\ell(t_1,t_2) +
\int_{{\bf R}^2} dt_3 dt_4 Z^k(t_2,t_3) \beta_\ell(t_3,t_4)
\delta^{\ell m}\beta_m(t_1,t_4).
\eeq
We see that the new drift term clearly is not acting tangential to 
the gauge orbit; its rather complicated structure is necessary for 
leaving unchanged gauge invariant expectation values. We prove 
that this is indeed so. 

Suppose that $F$ is a gauge invariant functional of the fields $\Phi^k$, 
so that
\beq
\int dt \frac{\delta F}{\delta \Phi^i(t)} Z^i(t,t') = 0.
\eeq
We consider now the difference of the expectation values of $F$
when  we first use the generalized Langevin 
equation (5.3) and second the original Parisi-Wu 
equation (5.1), which is just the $\alpha=0$, $\beta_k=0$ 
contribution of 
(5.3). With the Ito rule (2.7) we find straightforwardly
\beqa
\lefteqn{\left. \langle\langle dF\rangle\rangle - \langle\langle 
dF\rangle\rangle
\right|_{\alpha=0,\beta_k=0} =
\langle\langle \int_{{\bf R}^2} dt_1 dt_2
\left[ \frac{\delta F}{\delta \Phi^i(t_1)} Z^i(t_1,t_2) \alpha(t_2)\right.}
\no \\
&& \mbox{} + \left.\left. \int_{\bf R} dt_3 
\frac{\delta}{\delta \Phi^k(t_3)} 
\left( \frac{\delta F}{\delta \Phi^i(t_1)} Z^i(t_1,t_2)\right)
\zeta^k(t_2,t_3) \right] \right|_{\Phi(\cdot)=\Phi(\cdot,s)} \rangle\rangle ds
\eeqa
which is zero, as each term vanishes separately due the gauge 
invariance of $F$.

Our next task is to discuss the issue of stochastic gauge fixing in 
terms of the Fokker-Planck equation. Zwanziger originally worked 
in this scheme to derive his stochastic gauge fixing procedure; we 
will now generalize his formulation. 

A remarkable advantage of the Fokker-Planck scheme in 
comparison to the Langevin formulation is that equilibrium 
properties of the system may directly be studied in a 
nonperturbative way. By this we mean that one neither needs to 
explicitly find the solution of the stochastic differential equation 
of the fields $\Phi^k$, nor is it necessary to solve the Fokker-Planck 
equation for the probability distribution $\rho$ for all stochastic 
times $s$. What, however, one may directly be able to achieve is to 
extract from the Fokker-Planck equation just the equilibrium 
distribution. Surely one must be cautious to check that all 
conditions for the stochastic process really to relax to this 
equilibrium distribution are fulfilled. 

We will demonstrate below that our generalized stochastic gauge 
fixing formulation will indeed allow us to extract in a 
nonperturbative manner an equilibrium Fokker-Planck 
probability distribution which defines a standard field theory 
path integral density for the helix model. 

First we want to repeat within the Fokker-Planck approach that 
gauge invariant expectation values remain unchanged when our 
generalized stochastic gauge fixing procedure is applied. We have 
to study the stochastic time development of gauge invariant 
expectation values and therefore are interested in the form of the 
adjoint Fokker-Planck operator $L^\dg$. Corresponding to the Langevin 
equation (5.3) and following the general discussion (2.1) and 
(2.12)  we obtain
\beqa
L^\dg[\Phi] &=& \int_{\bf R} dt \left[ - \frac{\delta S}{\delta \Phi^i(t)}
+ \frac{\delta}{\delta \Phi^i(t)} \right]
\delta^{ij}  \frac{\delta}{\delta\Phi^j(t)} \no \\
&& \mbox{} + \int_{{\bf R}^2} dt_1 dt_2 Z^i(t_1,t_2) \alpha(t_2)
\frac{\delta}{\delta \Phi^i(t_1)} \no \\
&& \mbox{} + \int_{{\bf R}^3} dt_1 dt_2 dt_3 \zeta^k(t_2,t_3)
\frac{\delta}{\delta \Phi^k(t_3)}
\left( Z^i(t_1,t_2) \frac{\delta}{\delta \Phi^i(t_1)}\right).
\eeqa
We recognize that each of the last two terms is a
differential operator acting along the gauge orbits so that it vanishes 
when it is applied 
to gauge invariant observables. Thus we have demonstrated 
within the Fokker-Planck formulation as well that expectation 
values of gauge invariant observables are not affected by our 
modifications.

Comparing with Zwanziger's original stochastic gauge fixing 
procedure our generalized scheme amounts to adding to the 
(adjoint) Fokker-Planck operator not only a gauge symmetry 
preserving part which is linear in the field derivatives, but also a 
gauge symmetry preserving part which contains quadratic 
derivatives. In other words not only the drift term but also the 
diffusion term of the (adjoint) Fokker-Planck operator gets 
modified; the modification of the latter corresponds precisely to 
the modifcation of the Wiener process of the Langevin equation 
(5.3).

It is of crucial importance in Zwanziger's approach to draw 
advantage of the fact that the expectation values of gauge 
invariant quantities are independent of the new drift term, which 
in specific means that they are independent of the function $\alpha$. 
One  may chose now such a specific value for it that the functional 
integrations along the gauge orbits receive proper damping. 
Equivalently the gauge modes' contributions to expectation values 
are seen to converge rapidly, i.e. exponentially, in stochastic time. 
Only with such a proper choice the Fokker-Planck density can be 
normalized to unity and allows for a true probabilistic 
interpretation of the stochastic scheme. 

We want to discuss the issue of proper damping of the gauge 
modes within our generalized scheme in the next section.

\section{Damping of the gauge degrees of freedom}

So far we explained our stochastic gauge fixing method in quite 
general terms, but we now want to consider gauge models where 
the transition to gauge invariant variables at the one hand  
and gauge transforming variables on the other hand globally 
is possible. Geometrically this corresponds to a situation where the 
configuration space admits a structure of a trivialisable principal 
fiber bundle. The transition to these new variables is not unique 
and corresponds to a specific choice of gauge.

In the following we specialize further and discuss the helix model 
in detail.  In this way we can make use of our geometrical analysis 
in chapter 4) and concentrate on the two sets  (4.15, 4.17) of field 
variables $\Psi_{(\sigma_i)}$, $i = 1,2$, which are related to the gauge 
fixing 
surfaces 
$\Phi^3=0$ and $A=0$, respectively. We remind that in both cases the 
gauge dependent variables are changing under gauge 
transformations  just by a simple shift. 

We want to investigate how the  Langevin equations (5.3) are 
changed when we transform the original field variables $\Phi$ into 
one of the sets of new coordinates $\Psi_{(\sigma_i)}$. With respect to 
these 
variable changes we introduce the vielbeins $E$ and their inverses 
$e$, as well as the induced inverse metric $G^{\mu\nu}$; see (3.4-3.5), 
as well as the Appendix  for explicit expressions. 

With the use of the Ito formula (2.6) the Langevin equations (5.3) 
are transformed in a similar way as (3.7) into 
\beqa
d\Psi^\mu(t,s) &=&
\int_{\bf R} dt_1 \left\{ \left[ -G^{\mu\nu}(t,t_1)
\frac{\delta S}{\delta \Psi^\nu(t_1)}
+ \frac{\delta^2 \Psi^\mu(t)}{\delta \Phi^i(t_1)\delta \Phi^i(t_1)}
\right.\right. \no \\
&& \mbox{} + \left. \int_{\bf R} dt_2 E^\mu{}_i(t,t_1) Z^i(t_1,t_2)
\alpha(t_2) \right] ds \no \\
&& \mbox{} + \int_{\bf R} dt_2 E^\mu{}_i(t,t_1) \left[ 
\delta^i{}_k \delta(t_1-t_2) + \int_{\bf R} dt_3 Z^i(t_1,t_3)
\beta_k(t_3,t_2) \right] dW^k(t_2,s) \no \\
&& \mbox{} + \int_{{\bf R}^3} dt_2 dt_3 dt_4
\left.\left. \left[ \frac{\delta}{\delta \Phi^k(t_2)}
(E^\mu{}_i(t,t_4) Z^i(t_4,t_1))\right]
\zeta(t_1,t_2)\right\}\right|_{\Psi(\cdot) = \Psi(\cdot,s)}.
\eeqa
Fortunately the last term in this somewhat involved looking 
expression vanishes. This follows from  
 \beq
Z^i(t,t') = e^i{}_4(t,t')
\eeq
and the fact that $e$ is the 
inverse to $E$.
The expression (6.2) can  be derived by considering that the new 
coodinates 
$\Psi$ are 
gauge invariant except for $\Psi^4$ which, however, is transforming  only 
by a linear shift, see the discussion below (4.13). 
 Therefore under an infinitesimal gauge transformation $\xi \in 
\mbox{Lie }\G$ the fields $\Psi^\mu(t)$ transform according to the 
Lie-derivation 
along the vector field  $Z_\xi$,
\beqa
Z_\xi(\Psi^\mu(t)) &=& \int_{{\bf R}^2} dt_1 dt_2
\xi(t_1) Z^i(t_1,t_2)
\frac{\delta \Psi^\mu(t)}{\delta \Phi^i(t_1)}  \no \\
&=& \int_{{\bf R}^2} dt_1 dt_2 
\xi(t1) Z^i(t_1,t_2) E^\mu{}_i(t,t_1) 
\no \\
&=& \delta^\mu{}_4 \; \xi(t)
\eeqa
 so that (6.2) can be obtained. We also have
\beq
\int_{\bf R} dt \; \frac{\delta E^\nu{}_i(t,t')}{\delta \Psi^\nu(t)} = 0
\eeq
which is satisfied by both of the above sets of coordinates 
$\Psi_{(\sigma)}$
(see the Appendix) so that the Ito term takes the special form
\beqa
\int_{\bf R} dt_1 \frac{\delta^2 \Psi^\mu(t)}{\delta \Phi^i(t_1)
\delta \Phi^i(t_1)} &=&
\int_{{\bf R}^2} dt_1 dt_2 E^\nu{}_i(t_2,t_1)
\frac{\delta E^\mu(t,t_1)}{\delta \Psi^\nu(t_2)} \no \\
&=& \int_{\bf R} dt_2 \; \frac{\delta}{\delta \Psi^\nu(t_2)} \;
G^{\nu\mu}(t_2,t).
\eeqa
Thus we finally find the Langevin equations for the new variables        
$\Psi_{(\sigma)}$
\beqa
d\Psi^\mu(t,s) &=& \int_{\bf R} dt_1 \left\{ \left[ -G^{\mu\nu}(t,t_1)
\frac{\delta S}{\delta \Psi^\nu(t_1)} + 
\frac{\delta G^{\nu\mu}(t_1,t)}{\delta \Psi^\nu(t_1)} +
\delta^\mu{}_4 \delta(t-t_1) \alpha(t_1) \right] ds \right. \no \\
&& \mbox{} + \left.\left. [E^\mu{}_i(t,t_1) + \delta^\mu{}_4 
\beta_i(t,t_1)]
dW^i(t_1,s)\right\} \right|_{\Psi(\cdot) = \Psi(\cdot,s)}.
\eeqa
We recognize that the terms due to the generalized gauge fixing 
procedure - the terms containing the arbitrary functions $\alpha$ and 
$\beta_k$ - are not affecting the Langevin equations of the gauge 
independent fields $\Psi^{\bar\mu}_{(\sigma)}$, $\bar\mu = 1,2,3$
\beqa
d\Psi^{\bar\mu}(t,s) &=& \int_{\bf R} dt_1 \left\{ \left[ -G^{\bar{\mu}\bar\nu}
(t,t_1)\frac{\delta S}{\delta \Psi^{\bar\nu}(t_1)} + 
\frac{\delta G^{\bar{\nu}\bar\mu}(t_1,t)}{\delta \Psi^{\bar\nu}(t_1)} 
 \right] ds \right. \no \\
&& \mbox{} + \left.\left. E^{\bar\mu}{}_i(t,t_1) 
dW^i(t_1,s)\right\} \right|_{\Psi(\cdot) = \Psi(\cdot,s)}.
\eeqa
It is important to remark that these Langevin equations are closed among 
themselves and can be solved independently of the gauge 
dependent variable $\Psi^4$. This may look surprising, as the vielbein E 
generally is a gauge dependent matrix ($G^{\mu\nu}$, however,
is gauge 
independent by direct inspection).  The claimed gauge independence of the above Langevin 
equations  
  follows  because

\begin{enumerate}
\item[i)] the 
vielbein $E$ can be split into the product of a gauge invariant matrix and a pure rotation matrix which contains all the gauge dependence (see (A.3) and (A.7))

\item[ii)] the Ito-Langevin 
equations are invariant under rotations of the vielbeins (see 
\cite{Graham} and references therein), so that the gauge dependent part of $E$ can just  be rotated away.

\item[iii)] The gauge independence of
 $G^{\mu\nu}$  follows generally from (3.5) as  a direct 
consequence of the product form of $E$.  
\end{enumerate}

At this point it seems appropriate to notice that equation (6.7) can also 
 be obtained within the pure Parisi-Wu formulation of gauge theories, i.e.
  without introducing any stochastic gauge fixing terms. In this case the original Langevin equation (5.1) constitutes  the starting point for introducing the new coordinates $\Psi$,  all the steps concerning the application of the Ito formula are the same as above.
The formulation (6.7) in terms 
of only gauge invariant fields
 is precisely in the spirit of the original paper of Parisi-Wu \cite{Parisi+Wu} where the transition to gauge invariant variables and the discarding of the 
gauge variant variables was suggested. In our case we are, however, 
particulary interested in connecting the stochastic formalism to the path 
integral scheme. We therefore prefer to consider the stochastic time evolution of 
all the variables of the gauge model and continue within the generalized stochastic gauge fixing approach.

In order to derive the Langevin equation of the gauge dependent 
field $\Psi^4$ we use
\beq
\int dt' \; \frac{\delta G^{\nu 4}(t',t)}{\delta \Psi^\nu(t')} =
\int dt' \; \frac{\delta E^4{}_i(t,t')}{\delta \Phi^i(t')} = 0
\eeq
This follows from (6.4) and the fact that $\Psi^4$ is linear in $\Phi$, 
see (4.15)  and (4.17), as well as (4.6). We obtain
\beqa
d\Psi^4(t,s) &=& \int_{\bf R} dt_1 \left\{ \left[ -G^{4\nu}(t,t_1)
\frac{\delta S}{\delta \Psi^\nu(t_1)} + \alpha(t_1) \delta(t-t_1)
\right] ds \right. \no \\
&& \mbox{} + \left.\left. [E^4{}_k(t,t_1) + \beta_k(t,t_1)]dW^k
(t_1,s) \right\} \right|_{\Psi(\cdot) = \Psi(\cdot,s)}.
\eeqa
We note that on the right hand side of (6.9)  no gauge dependent 
terms are present apart of the possibility that the so far arbitrary 
functions $\alpha$ and $\beta_k$ are changing under gauge 
transformations. Let us suppose, however, for a moment that 
$\alpha$ is gauge invariant. In this case the drift term of (6.9) is 
gauge invariant as well and remains unchanged for a stochastic 
evolution along the gauge orbits, i.e. along the gauge transforming 
variable $\Psi^4$ keeping the other variables fixed. For this 
stochastic process the drift term of the Langevin equation of 
$\Psi^4$ appears like a constant; it gives rise to a diffusion 
behaviour and will never allow relaxation to an equilibrium 
distribution.

But we can convert this diffusion into a well damped stochastic 
process by conveniently choosing a gauge dependent $\alpha$: We 
recall that due to the general philosophy of stochastic gauge fixing, 
which we have formulated in the previous section, every choice of 
$\alpha$ (as well as of $\beta_k$) is permissible. 

Out of the possible choices the simplest Langevin equation  is obtained 
if $\alpha$ subtracts off the above gauge invariant drift term and 
adds a term proportional to $\Psi^4$ itself, the proportionality factor 
being built from a gauge invariant and strictly positive integral kernel 
$\gamma$ 
\beq
\alpha(t) = \int_{\bf R} dt' \left[ G^{4\nu}(t,t')
\frac{\delta S}{\delta \Psi^\nu(t')} -
\gamma(t,t') \Psi^4(t')\right].
\eeq
We can suggestively rewrite this damping term by introducing the 
total action $S_{\rm tot}$
\beq
S_{\rm tot} = S + \frac{1}{2} \int_{\bf R} dt
(\Psi^4(t))^2
\eeq
so that the $\Psi^4$ Langevin equation becomes
\beqa
d\Psi^4(t,s) &=& \int_{\bf R} dt_1 \left\{ -\gamma(t,t_1)
\frac{\delta S_{\rm tot}}{\delta \Psi^4(t_1)} \right. \no \\
&& \mbox{} + \left.\left. [E^4{}_i(t,t_1) + \beta_i(t,t_1)]dW^i
(t_1,s) \right\} \right|_{\Psi(\cdot) = \Psi(\cdot,s)}.
\eeqa
Let us define a  matrix $\wt G$ by
\beqa
\wt G^{\bar\mu \bar\nu}(t_1,t_2) &=& G^{\bar\mu \bar\nu}(t_1,t_2)
\qquad \bar\mu, \bar\nu = 1,2,3 \no \\
\wt G^{\bar\mu 4}(t_1,t_2) &=& \wt G^{4 \bar\mu}(t_1,t_2) = 0 \no \\
\wt G^{44}(t_1,t_2) &=& \gamma(t_1,t_2).
\eeqa
Notice that at this moment it is not  obvious that 
$\wt G$ as defined above is a positive matrix and therefore
 satisfies the properties of a metric; 
we will prove it, however, in the next section for a special choice of
$\beta_k$ and $\gamma$. Since from (6.8)
\beq
\frac{\delta \wt G^{\mu\nu}(t_1,t_2)}{\delta \Psi^\nu(t_2)} =
\frac{\delta G^{\mu\nu}(t_1,t_2)}{\delta \Psi^\nu(t_2)}
\eeq
 the Langevin equations 
(6.6) can be rewritten in terms of the total action $S_{\rm tot}$
and the new metric $\wt G$ as follows
\beqa
d\Psi^\mu(t,s) &=& \int_{\bf R} dt_1 \left\{ \left[ - \wt G^{\mu\nu}
(t,t_1) \frac{\delta S_{\rm tot}}{\delta \Psi^\nu(t_1)} +
\frac{\delta \wt G^{\nu\mu}(t_1,t)}{\delta \Psi^\nu(t_1)}
\right] ds \right. \no \\
&& \mbox{} + \left.\left. \left[ E^\mu{}_k(t,t_1) + \delta^\mu{}_4
\beta_k(t,t_1)\right] dW^k(t_1,s) \right\} \right|_{\Psi(\cdot) =
\Psi(\cdot,s)}.
\eeqa
Before closing this section we rewrite the adjoint Fokker-Planck 
operator $L^\dg$ in the new coordinates $\Psi_{(\sigma)}$. 
From (6.6) we get easily
\beqa
L^\dg[\Psi] &=& \int_{\bf R} dt \left\{ \int_{\bf R} dt_1 
\left[ - \frac{\delta S}{\delta \Psi^\nu(t_1)} +
\frac{\delta}{\delta \Psi^\nu(t_1)}\right]
G^{\mu\nu}(t_1,t) \frac{\delta}{\delta \Psi^\mu(t)}
+ \alpha(t) \frac{\delta}{\delta \Psi^4(t)} \right. \no \\
&& \mbox{} + \left. \int_{{\bf R}^2} dt_1 dt_2 \beta_k \delta^{k\ell}
\left[ 2E^\nu{}_\ell(t_1,t_2) 
\frac{\delta^2}{\delta \Psi^\nu(t_1) \delta \Psi^4(t)} +
\beta_\ell(t_1,t_2)
\frac{\delta^2}{\delta \Psi^4(t_1) \delta \Psi^4(t)} \right] \right\}.
\no \\
\eeqa
It is evident that the additional terms due to our modifications 
exhaust the most general structure of the adjoint Fokker-Planck 
operator compatible with gauge invariance.

We finally rewrite $L^\dg$ upon inserting the special value (6.10) of 
$\alpha$ as
\beqa
L^\dg[\Psi] &=& \int_{{\bf R}^2} dt dt_1 \left\{ \left[ 
- \frac{\delta S_{\rm tot}}{\delta \Psi^\nu(t_1)} +
\frac{\delta}{\delta \Psi^\nu(t_1)}\right]
\wt G^{\nu\mu}(t_1,t) \frac{\delta}{\delta \Psi^\mu(t)} \right. \no \\
&& \mbox{} + \left[ G^{\mu\nu}(t,t_1) - \wt G^{\mu\nu}(t,t_1)\right]
\frac{\delta^2}{\delta \Psi^\nu(t_1) \delta \Psi^\mu(t)} \no \\
&& \mbox{} + \left. \int_{\bf R} dt_2 \beta_k \delta^{k\ell}
\left[ 2E^\nu{}_\ell(t_1,t_2) 
\frac{\delta^2}{\delta \Psi^\nu(t_1) \delta \Psi^4(t)} +
\beta_\ell(t_1,t_2)
\frac{\delta^2}{\delta \Psi^4(t_1) \delta \Psi^4(t)} \right] \right\}.
\no \\
\eeqa
In retrospect the use of the new coordinates $\Psi_{(\sigma)}$ has turned 
out 
to be successful in order to analyze the damping behaviour of the 
stochastic process.  It is 
explicitly shown that the process along the gauge group, although 
initially giving rise to pure diffusion, can be damped by an 
appropriate tuning of the gauge fixing parameter $\alpha$.

\section{The equilibrium distribution}

In this section we exploit the remaining freedom for the choice of 
the positive kernel $\gamma$ and the vector $\beta_k$ to define a 
specific stochastic process of which we are able to derive the 
equilibrium distribution. 

We repeat once more that the legitimation for the freedom of 
choosing $\gamma$ and $\beta_k$ was given in section 5 where we 
proved that the expectation values of gauge invariant observables 
are independent of $\gamma$ and $\beta_k$. Therefore, if we are able 
to find just one stochastic process - defined in terms of specific 
functions $\alpha$ and $\beta_k$ - where the equilibrium distribution is 
known to us, we have succeeded in finding a standard path 
integral density leading to unambiguous expectation values for 
gauge invariant quantitites. Eventually we can compare it with the 
path integral density derived by conventional techniques, to name 
the Faddeev-Popov procedure. 

The idea for our procedure is to start from the class of well 
converging stochastic processes (6.15) and choose in a self-consistent way 
such specific $\beta_k$  and $\gamma$ that the kernel 
$\wt G$ appears in the Fokker-Planck operators in just factorized 
form. This corresponds to a situation where the 
fluctuation-dissipation theorem applies, which guarantees that in the 
equilibrium limit the stochastic process becomes independent of 
$\wt G$. We will see that the equilibrium distribution of our 
specially choosen stochastic process is proportional to the 
exponent of the  gauge fixed total action $S_{\rm tot}$.

To become more specific we recall (6.17), where the adjoint 
Fokker-Planck operator $L^\dg$ was written out in detail. Suppose that 
we find a solution for $\beta_k$  and $\gamma$ such that
\beq
\wt G^{\mu\nu}(t_1,t_2) = \int_{\bf R} dt_3 [E^\mu{}_k(t_1,t_3) +
\delta^\mu{}_4 \beta_k(t_1,t_3)] \delta^{k\ell}[E^\nu{}_\ell(t_2,t_3)
+ \delta^\nu{}_4 \beta_\ell(t_2,t_3)]
\eeq
then indeed $\wt G$ is appearing just in factorized form and we can 
simplify drastically the adjoint Fokker-Planck operator
\beq
L^\dg[\Psi] = \int_{{\bf R}^2} dt_1 dt_2 
\left[ - \frac{\delta S_{\rm tot}}{\delta \Psi^\nu(t_1)} +
\frac{\delta}{\delta \Psi^\nu(t_1)} \right]
\wt G^{\nu\mu}(t_1,t_2) \frac{\delta}{\delta \Psi^\mu(t_2)}
\eeq
as well as the Fokker-Planck operator itself
\beq
L[\Psi] = \int_{{\bf R}^2} dt_1 dt_2 \frac{\delta}{\delta \Psi^\mu(t_1)}
\wt G^{\mu\nu} \left[ \frac{\delta S_{\rm tot}}{\delta \Psi^\nu(t_2)}
+ \frac{\delta}{\delta \Psi^\nu(t_2)} \right].
\eeq
It is not difficult to give a solution for $\beta_k$ and $\gamma$ which 
fulfills (7.1). First we deduce  from (6.13) the conditions
\beq
\int_{\bf R} dt_3 E^{\bar\mu}{}_k(t_1,t_3) \delta^{k\ell}
\left[ E^4{}_\ell(t_2,t_3) + \beta_\ell(t_2,t_3)\right] = 0 \qquad
\bar \mu = 1,2,3
\eeq
as well as
\beq
\int_{\bf R} dt_3 \left[E^4{}_k(t_1,t_3) + \beta_k(t_1,t_3)\right]
\delta^{k\ell} \left[E^4{}_\ell(t_2,t_3) + \beta_\ell(t_2,t_3)\right]
= \gamma(t_1,t_2)
\eeq
and easily find the self-consistent solutions 
\beq
\beta_k(t_1,t_2) = - E^4{}_k(t_1,t_2) + \delta_{k\ell} e^\ell{}_4
(t_2,t_1)
\eeq
and
\beq
\gamma(t_1,t_2) = \int_{\bf R} dt_3 e^k{}_4(t_3,t_1) \delta_{k\ell}
e^\ell{}_4(t_3,t_2).
\eeq
In the last section we claimed that $\wt G$ could be interpreted as a 
metric. From (7.1) and from (7.6-7.7) we see that this is indeed 
the case: We introduce the vielbein $\wt E$
\beq
\wt E^\mu{}_k(t_1,t_2) = E^\mu{}_k(t_1,t_2) + \delta^\mu{}_4 
[- E^4{}_k(t_1,t_2) + \delta_{k\ell} e^\ell{}_4
(t_2,t_1)]
\eeq
with the help of which $\wt G$ is explicitly decomposable as
\beq
\wt G^{\mu\nu}(t_1,t_2) = \int_{\bf R} dt_3 \wt E^\mu{}_k(t_1,t_3)
\delta^{k\ell} \wt E^\nu{}_\ell(t_2,t_3)
\eeq
and by construction is a positive matrix. 

We would like to remark, however, that the vielbeins $\wt E$ are 
anholonomic as in specific
\beq
\frac{\delta \wt E^\mu{}_i(t_1,t_2)}{\delta \Phi^k(t_3)} -
\frac{\delta \wt E^\mu{}_k(t_1,t_2)}{\delta \Phi^i(t_3)} =
\delta^\mu{}_4 \left[ \delta_{i\ell}
\frac{\delta e^\ell{}_4(t_2,t_1)}{\delta \Phi^k(t_3)} -
\delta_{k\ell}
\frac{\delta e^\ell{}_4(t_2,t_1)}{\delta \Phi^i(t_3)} \right] \neq 0
\eeq
so that there does not exist a coordinate sytem with respect to 
which the vielbeins $\wt E$ could be interpreted as the jacobians.

After this discussion of the self-consistent solutions (7.6-7) for 
$\gamma$ and $\beta_k$ which lead to factorized Fokker-Planck 
operators we now are interested in writing down the corresponding 
Langevin equations. They read
\beqa
d\Psi^\mu(t,s) &=& \int_{\bf R} dt_1 \left\{ \left[ -\wt G^{\mu\nu}
(t_,t_1) \frac{\delta S_{\rm tot}}{\delta \Psi^\nu(t_1)} +
\frac{\delta \wt G^{\nu\mu}(t,t_1)}{\delta \Psi^\nu(t_1)} \right]ds
\right. \no \\
&& \mbox{} + \left.\left. \wt E^\mu{}_k(t,t_1) dW^k(t_1,s)\right\}
\right|_{\Psi(\cdot) = \Psi(\cdot,s)}
\eeqa
and are of a similar form as discussed at the end of section 3). Like in 
the case of (6.7) the vielbein $\wt E$ factorizes in a gauge invariant 
part and a gauge dependent rotation matrix (which could in principle be 
absorbed, due to the previously mentioned properties of Ito-Langevin 
equations). 

We now derive the equilibrium distribution of the stochastic 
process described by the above Langevin equation or by the 
Fokker-Planck equation with Fokker-Planck operator given by 
(7.3). We remind that we restricted ourselves to the class of well 
converging stochastic processes, see (6.15) as well as (7.6-6.7), so 
that the Fokker-Planck probability distribution is normalizable. 
Most crucially we have that $\wt G$ is positive, see (7.9); it is 
appearing in the Fokker-Planck operator in factorized form, see 
(7.3). As a consequence the formal stationary limit of the Fokker-
Planck probability distribution can be identified with the 
equilibrium limit and reads
\beq
\rho[\Psi]_{\rm equil.} = \frac{e^{-S_{\rm tot}}} {\int D \Psi
e^{- S_{\rm tot}}}.
\eeq
Now we use the explicit forms of the total actions
to explicitly present our result for the path integral 
density. 

In this paper we discussed two specific gauge fixing surfaces; for 
the case of the variables $\Psi_{(\sigma_1)}$ corresponding to the gauge 
fixing 
surface $\vp^3=0$, after integrating out $\Psi^3_{(\sigma_1)}$ and 
$\Psi^4_{(\sigma_1)}$, our path 
integral density obtains
\beq
\det \left(1 + (\Psi^1_{(\sigma_1)})^2 +
( \Psi^2_{(\sigma_1)})^2 \right)^{-1/2} 
e^{-S_{\rm eff} [\Psi^1_{(\sigma_1)},\Psi^2_{(\sigma_1)}]}.
\eeq
Here the effective action in terms of $\Psi^1$ and $\Psi^2$ is given by
\beqa
\lefteqn{S_{\rm eff}[
\Psi^1_{(\sigma_1)},\Psi^2_{(\sigma_1)}
] = \int dt \frac{1}{2} \cdot } \no \\
&&[ (\dot \Psi^1_{(\sigma_1)},\dot \Psi^2_{(\sigma_1)}) 
\left( \ba{cc}
1 - \dfrac{(\Psi^2_{(\sigma_1)})^2}
{1 + (\Psi^1_{(\sigma_1)})^2 + (\Psi^2_{(\sigma_1)})^2} &
\dfrac{\Psi^1_{(\sigma_1)} \Psi^2_{(\sigma_1)}}
{1 + (\Psi^1_{(\sigma_1)})^2 + (\Psi^2_{(\sigma_1)})^2}  \\[12pt]
\dfrac{\Psi^1_{(\sigma_1)} \Psi^2_{(\sigma_1)}}
{1 + (\Psi^1_{(\sigma_1)})^2 + (\Psi^2_{(\sigma_1)})^2}  &
1 - \dfrac{(\Psi^1_{(\sigma_1)})^2}
{1 + (\Psi^1_{(\sigma_1)})^2 + (\Psi^2_{(\sigma_1)})^2} \ea \right)
\left( \ba{c}
\dot \Psi^1_{(\sigma_1)} \\ \dot \Psi^2_{(\sigma_1)} \ea \right)
\no \\
&& \mbox{} + (\Psi^1_{(\sigma_1)})^2 + (\Psi^2_{(\sigma_1)})^2]
\eeqa
and is agreeing nicely with the result of \cite{Friedberg et al.}.

For the coordinates $\Psi_{(\sigma_2)}$ 
 the integration over $\Psi^3_{(\sigma_2)}$ 
and $\Psi^4_{(\sigma_2)}$  does not leave any 
trace and the path integral measure is just the exponent of the 
extremely simple effective action
\beq
S_{\rm eff}[\Psi^1_{(\sigma_2)},\Psi^2_{(\sigma_2)}] = 
\int dt \frac{1}{2} \left[ 
(\dot \Psi^1_{(\sigma_2)}(t))^2 + (\dot \Psi^2_{(\sigma_2)}(t))^2 
+ (\Psi^1_{(\sigma_2)}(t))^2 + (\Psi^2_{(\sigma_2)}(t))^2
\right]. 
\eeq
It is evident that this gauge fixing leads to a quadratic action and 
it is therefore completely trivial to calculate the only relevant 
gauge invariant quantity, the  two point function
\beq
\lim_{s \ra \infty} \langle (\Psi^1_{(\sigma_2)}(t))^2 +
(\Psi^2_{(\sigma_2)}(t))^2\rangle (s) = 2\pi .
\eeq
It nicely agrees with the corresponding result in Dirac's scheme 
(see e.g. \cite{Friedberg et al.} and \cite{Fujikawa}) or with the path 
integral results in \cite{Friedberg et al.}.

At the end of this section we transform the final Langevin 
equation (7.11)  back into the 
original cartesian coordinates $\Phi^k$. Amusingly we end up with a 
Langevin equation with a similar structure as in (7.11): First we 
have with the Ito rules
\beqa
d\Phi^i(t,s) &=& \int_{\bf R} dt_1 \left[ e^i{}_\mu(t,t_1) 
d \Psi^\mu(t_1) \right.  \no \\
&& \mbox{} + \left. \left. \int_{\bf R} dt_2 
\frac{\delta e^i{}_\mu(t,t_1)}{\delta \Psi^\nu(t_2)}
\wt G^{\mu\nu}(t_1,t_2) \right] \right|_{\Psi(\cdot) = \Psi(\cdot,s)}.
\eeqa
Due to the fact that the $\Phi$ are flat coordinates we have - similarly 
 as (3.9) -
\beq
\int dt_1 \; \frac{\delta e^k{}_\mu(t_1,t_2)}{\delta \Phi^k(t_1)}
= 0
\eeq
so that we finally arrive at
\beqa
d\Phi^i(t,s) &=& \int_{\bf R} dt_1 \left\{ \left[
- \hat g^{ik}(t,t_1) \frac{\delta S_{\rm tot}}{\delta \Phi^k(t_1)}
+ \frac{\delta \hat g^{ki}(t_1,t)}{\delta \Phi^k(t_1)} \right] ds
\right. \no \\
&& \mbox{} + \left.\left. \hat e^i{}_k(t,t_1) dW^k(t_1,s) \right\}
\right|_{\Phi(\cdot) = \Phi(\cdot,s)}.
\eeqa
Here we defined new vielbeins $\hat e$
\beq
\hat e^i{}_k(t_1,t_2) = \delta^i{}_k \delta(t_1-t_2) + \int_{\bf R} dt_3
e^i{}_4(t_1,t_3) [ 
- E^4{}_k(t_3,t_2) + \delta_{k\ell} e^\ell{}_4(t_2,t_3)]
\eeq
and a metric $\hat g$
\beq
\hat g^{ik}(t_1,t_2) = \int_{\bf R} dt_3 \; \hat e^i{}_\ell(t_1,t_3)
\delta^{\ell m} \hat e^k{}_m(t_2,t_3).
\eeq
We recognize that the Langevin equation (7.20) indeed has the 
same form as (7.11); the same conclusions apply about the 
application of the fluctuation dissipation theorem  and the 
independence of the stochastic process of $\hat g$ in the equilibrium 
limit.

We close with the remark that although the Langevin equation 
(7.20) is defined with respect to the original cartesian field 
coordinates the procedure of stochastic gauge fixing has 
introduced a dependence of the new coordinates $\Psi_{(\sigma)}$. 

\section{Geometrical aspects of the  generalized stochastic gauge fixing procedure}
The main
virtue of the generalized stochastic gauge fixing procedure lies in the fact
 that for specific choices of the extra functions $\alpha$ and $\beta_k$ the
fluctuation-dissipation theorem can be applied; as a consequence
the equilibrium distribution can be derived quite easily. 
Thereby we were led to introduce in (6.13)  a positive definite, matrix-valued
kernel $\wt G^{\mu\nu}$ which can be regarded as the
inverse of a particular matrix, denoted by $\wt G_{\mu\nu}$, on
$\Sigma_{(\sigma)} \times \G$. What can now be said about its geometrical
meaning?

We will begin with the explicit calculation of $\wt G_{\mu\nu}$.
First it can be shown by a direct calculation that the inverse vielbein of $\wt E^\mu{}_k$ which we denote by $\wt e^i{}_\mu$ is given by
\beqa
\wt e^i{}_\mu(t,t') &=& e^i{}_\mu(t,t') +
\int_{{\bf R}^3} dt_1 dt_2 dt_3 Z^i(t,t_1) K^{-1}(t_1,t_2) \cdot \no 
\\
&& [E^4_j(t_2,t_3) - \delta_{jk} Z^k(t_3,t_2)] e^j{}_\mu(t_3,t').
\eeqa
Here the kernel $K^{-1} \in C^\infty({\bf R}^2,{\bf R})$ is uniquely 
defined by
\beq
\int_{\bf R} dt' K^{-1}(t_1,t') G_{44}(t',t_2) = \delta(t_1 - t_2),
\eeq
where $G_{\mu\nu}$ denotes the metric inverse to $G^{\mu\nu}$,
\beq
G_{\mu\nu}(\tau_1,\tau_2) = \int_{\bf R} dt  e^i{}_\mu(t,\tau_1)
\delta_{ij}  e^j{}_\nu(t,\tau_2).
\eeq
and the following identities hold
\beqa
\int_{\bf R} dt \wt e^k{}_\mu(t_1,t) \wt E^\mu{}_m(t,t_2) &=&
\delta^k{}_m\delta(t_1 - t_2) \no \\
\int_{\bf R} dt \wt e^k{}_\mu(t_1,t) \wt E^\nu{}_k(t,t_2) &=&
\delta_\mu{}^\nu\delta(t_1 - t_2) .
\eeqa
The components of the metric $\wt G = \int_{{\bf R}^2} d\tau_1 d\tau_2 \wt G_{\mu\nu}
(\tau_1,\tau_2) \delta \Psi^\mu(\tau_1) \otimes \delta \Psi^\nu(\tau_2)$
are then given by
\beqa
\wt G_{\mu\nu}(\tau_1,\tau_2) &=& \int_{\bf R} dt \wt e^i{}_\mu(t,\tau_1)
\delta_{ij} \wt e^j{}_\nu(t,\tau_2) \no \\
&=& \int_{{\bf R}^2} d\tau d\tau' \int_{\bf R} dt \wt e^i{}_\mu(t,\tau_1)
E^{\lambda_1}{}_i(\tau,t) G_{\lambda_1\lambda_2}(\tau,\tau')
\int_{\bf R} dt' E^{\lambda_2}{}_j(\tau',t') \wt e^j{}_\nu(t',\tau_2) \no \\
&=& \wh G_{\mu\nu}(\tau_1,\tau_2) + \delta_{\mu}{}^4 \delta^4{}_{\nu} 
K^{-1}(\tau_1,\tau_2).
\eeqa
In the above equation we introduced
\beq
\wh G_{\mu\nu}(\tau_1,\tau_2) = G_{\mu\nu}(\tau_1,\tau_2) -
\int_{{\bf R}^2} dt dt' G_{\mu 4}(\tau_1,t) K^{-1}(t,t')
G_{4 \nu}(t',\tau_2)
\eeq

From a more
geometrical viewpoint, the matrix valued kernel
$\int_{\bf R} dt  E^\nu{}_i(\tau_1,t) \wt e^i{}_\mu(t,\tau_2)$ in
(8.5) can be regarded as component of a vector valued one form $\ve$
on $\Sigma_{(\sigma)} \times \G$, namely
\beq
\ve = \int_{{\bf R}^3} dt_1 dt_2 dt_3 E^\nu{}_i(t_1,t_2) 
\wt e^i{}_\mu(t_2,t_3) \delta\Psi^\mu(t_3) \otimes
\frac{\delta}{\delta \Psi^\nu(t_1)}.
\eeq
Insertion of (8.1) yields
\beqa
\ve &=& \int_{\bf R} dt \delta \Psi^\mu(t) \otimes
\frac{\delta}{\delta \Psi^\mu(t)} -
\int_{{\bf R}^3} dt_1 dt_2 dt_3 K^{-1}(t_1,t_2) G_{4\mu}(t_2,t_3)
\delta \Psi^\mu(t_3) \otimes \frac{\delta}{\delta \Psi^4(t_1)} \no \\
&& \mbox{} + \int_{{\bf R}^2} dt_1 dt_2 K^{-1}(t_1,t_2) \delta \Psi^4(t_2)
\otimes \frac{\delta}{\delta \Psi^4(t_1)}
\eeqa
so that for each vector field $X$ on $\Sigma_{(\sigma)} \times \G$ we
finally get the formula
\beq
\ve(X) = X - Z_{B(X)} + Z_{\Theta(X)}. 
\eeq
Here both
\beq
B(\tau) = \int_{{\bf R}^2} dt_1 dt_2 K^{-1}(\tau,t_1) G_{4\mu}(t_1,t_2)
\delta \Psi^\mu(t_2)
\eeq
and
\beq
\Theta(\tau) = \int_{\bf R} dt_1 K^{-1}(\tau,t_1) \delta \Psi^4(t_1)
\eeq
are Lie~$\G$-valued one forms on $\Sigma_{(\sigma)} \times \G$.
Since the form $B$ is $\G$-invariant and $B(Z_\xi) = \xi$ $\forall \;
\xi \in$~Lie~$\G$, it defines a connection in the trivial principal
$\G$-bundle $\Sigma_{(\sigma)} \times \G \ra \Sigma_{(\sigma)}$.
However, the corresponding horizontal subbundle, which is given by all
those vector fields $X$ on $\Sigma_{(\sigma)} \times \G$ which are
annihilated by $B$, i.e. $B(X) = 0$, is orthogonal to the
gauge orbits with respect to the original metric $G_{\mu\nu}$.
This can be proven by a direct calculation.

Hence the term $X - Z_{B(X)}$ in (8.9) precisely gives the 
projection of $X$ onto this horizontal subbundle. In summary, the
global form of $\wt G$ is given by
\beq
\wt G(X,Y) = G(\ve(X),\ve(Y)) = G(X - Z_{B(X)},Y - Z_{B(Y)}) +
G(Z_{\Theta(X)},Z_{\Theta(Y)})
\eeq
for any two vector fields $X,Y$ on $\Sigma_{(\sigma)} \times \G$. From our
discussion it is evident that the terms appearing in the second part
of equation (8.12) correspond exactly to $\wh G_{\mu\nu}(\tau_1,\tau_2)$
and $\delta_{\mu 4} \delta_{\nu 4} K^{-1}(\tau_1,\tau_2)$ of (8.4)
respectively. Hence $\wh G_{\mu\nu}$ is the induced metric on the
horizontal subbundle, which descends to a well-defined metric on
$\Sigma_{(\sigma)}$. In fact, for $\bar \mu, \bar \nu = 1,2,3$ this horizontal metric $\wh G_{\bar\mu \bar\nu}$  can be shown to  be the inverse of the kernel $G^{\bar\mu \bar\nu}$ appearing in the Langevin equations (6.7) of the gauge invariant variables.

Before going on, however, we recall that due to the result (7.10),
no coordinate system exists such that $\ve$ could be interpreted as a
jacobian.

The curvature $\F$ of the connection is given by
\beq
\F = \frac{1}{2} \int_{{\bf R}^2}dt_1 dt_2 \left[ \frac{\delta B_\mu(t_1)}
{\delta \Psi^\nu(t_2)} - \frac{\delta B_\nu(t_2)}{\delta \Psi^\mu
(t_1)} \right] \delta \Psi^\mu(t_1) \wedge \delta \Psi^\nu(t_2).
\eeq
An explicit calculation yields that it is nonvanishing. Hence there
does not exist a manifold whose tangent bundle would coincide with the 
horizontal subbundle defined by the connection $B$.

At the end we would like to give some comments on the geometrical meaning 
of (7.14) and (7.15). 
The action $\R$ in 
(4.2) induces a free action  $\hat{\R}$ on $\E$
\beq
\hat{\R}(\vec \vp,g)(t) = (R(g(t)) \Un{\vp}(t), \vp^3(t) - g(t)),
\eeq
in a natural way. As a consequence, the projection 
$\hat{\pi} : \E   \ra  \M:= \E /\G$, with  $\hat{\pi}(\vec \vp) =
[\vec \vp]_\M$, where $[\vec \vp]_\M$ denotes the equivalence class of
$\vec \vp$ with respect to the action $\hat{\R}$, also admits the structure
of a principal $\G$-bundle over $\M$, where the quotient is taken with 
respect to $\hat{\R}$. Moreover, $\E \times_\G \; \A$ can be regarded as 
an 
associated fiber bundle over $\M$ with typical fiber $\A$ and projection 
$\tilde{\pi}(\vec \vp,A) = \hat{\pi}(\vec \vp) =
[\vec \vp]_\M$. The section $\sigma_1$ which was defined in (4.4) induces 
a global section 
$\hat{\sigma }_1$ of $\hat{\pi} : \E   \ra  \M$ by 
$
\hat{\sigma}_1([\vec \vp]_\M) = (R(\vp^3)\Un{\vp},0)
$. Let  $\hat{\Sigma}_{(\sigma_1)} = \mbox{im } \hat{\sigma}_1$ denote the 
corresponding gauge fixing surface
in $\E$, then the projection  $\Sigma_{(\sigma_1)} \ra 
\hat{\Sigma}_{(\sigma_1)}$ yields a 
fiber bundle with typical fiber $\A$ equivalent to 
$\tilde{\pi}: \E \times_\G \; \A
\ra \M$. Integrating out $\Psi^4_{(\sigma_1)}$ corresponds to fiber 
integration along the gauge group in the principal bundle  $\E \times \A 
\ra \Sigma_{(\sigma_1)}$ and integrating out $\Psi^3_{(\sigma_1)}$ 
corresponds 
to fiber integration along the fiber $\A$ in  $\Sigma_{(\sigma_1)} 
\ra \hat{\Sigma}_{(\sigma_1)}$.

The geometrical interpretation of (7.15) is slightly different. The 
action $\R$ in (4.2) also yields an induced action of $\G$ on $\A$, which 
although it is not free, also \cite{Bredon,Janich} gives rise to a smooth 
quotient $\N:=\A/\G$. Since every gauge potential $A \in \A$ can be 
gauged away by a suitable gauge transformation, the orbit space $\N$ 
consists only of the point $[0]$. Like in the former case,  $\E \times_\G 
\; \A$ can be regarded as a fiber bundle over $\N=\{[0]\}$ with typical 
fiber $\E$. Integrating out $\Psi^4_{(\sigma_2)}$ now corresponds to 
fiber integration along $\G$ in the principal bundle $\E \times \A 
\ra \Sigma_{(\sigma_2)}$. Since $\Sigma_{(\sigma_2)}\simeq\{0\} \times\E$, 
integrating 
out $\Psi^i_{(\sigma_2)}$, $i= 1,2,3$, corresponds purely to an 
integration over $\E$.

\section{Outlook}

In this paper we performed the stochastic quantization of the helix model  and were able to show agreement with the path integral procedure in a nonperturbative way. 

We generalized the stochastic gauge fixing procedure by  not only introducing an extra drift term horizontally to the gauge orbits, but by also modifying the Wiener process itself. This modification was done in such a way that  expectation values of gauge invariant objects remained unchanged. We gave a detailed analysis of the underlying geometrical structures of the model. Although the newly introduced extra terms at first sight seemed to render our approach  complicated, just the contrary  was proven to be the case: Due to a delicate interplay between the total drift term and the modified  diffusion term  the fluctuation dissipation theorem could be applied and the equilibrium limit  of the stochastic process was  derivable in a straightforward manner.

As one out of several immediate applications of our nonperturbative stochastic approach we are presently studying the helix model for the case where gauge fixing surfaces can be defined only locally; another work in progress is the phase space formulation of the helix model and the introduction of the BRST scheme. 

\section*{Acknowledgement}

We thank H. Grosse, H. Nakazato and K. Okano for valuable discussions. H. H. is grateful for the kind hospitality offered to him by the High Energy Physics Group of Tokuyama University, Tokuyama and by the High Energy Physics Group of the Dept. of Physics  of  Waseda University, Tokyo, where  this paper was completed; he is thankful for financial support from the Japanese Society of the Promotion of Science.

\appendix
\newcounter{zahler}
\renewcommand{\thesection}{\Alph{zahler}}
\renewcommand{\theequation}{\Alph{zahler}.\arabic{equation}}
\setcounter{zahler}{1}
\setcounter{equation}{0}
\section*{Appendix}
For the case of the variables $\Psi_{(\sigma_1)}$ (see (4.15) -- (4.16))
the vielbeins $E$ and $e$ (see (3.4)) are given by, respectively
\beq
E^\mu{}_i(t,t') = \left( \ba{cccc}
\cos \vp^3(t) & - \sin \vp^3(t) & - \Psi^2(t) & 0 \\
\sin \vp^3(t) &   \cos \vp^3(t) &   \Psi^1(t) & 0 \\
0            &           0      & - \partial_t & 1 \\
0            &           0      &   -1        &  0   
\ea \right) \delta(t-t')
\eeq
\beq
e^i{}_\mu(t,t') = \left( \ba{cccc}
\cos \vp^3(t) &  \sin \vp^3(t) &  0 & - \vp^2(t) \\
- \sin \vp^3(t) &   \cos \vp^3(t) & 0 & \vp^1(t) \\
0            &           0      & 0  & -1 \\
0            &           0      &   1        &  -\partial_t    
\ea \right) \delta(t-t')
\eeq
Furthermore we have
\beq
E^\mu{}_i(t,t') = \left( \ba{cccc}
1 & 0 & - \Psi^2(t) & 0 \\
0 & 1 & \Psi^1(t)   & 0 \\
0 & 0 & -\partial_t & 1 \\
0 & 0 & -1 & 0
\ea \right) 
\left( \ba{cccc}
\cos \vp^3(t) & - \sin \vp^3(t) & 0 & 0 \\
\sin \vp^3(t) &   \cos \vp^3(t) &   0 & 0 \\
0            &           0      & 1 & 0 \\
0            &           0      &   0 &  1   
\ea \right) \delta(t-t')
\eeq
In the case of the variables $\Psi_{(\sigma_2)}$ (see (4.17) -- (4.18), 
(4.6))
we have
\beq
E^\mu{}_i(t,t') = \left( \ba{cccc}
\cos F(t)\delta(t-t') & -\sin F(t) \delta(t-t') & -\Psi^2(t) h(t') & 
- \Psi^2(t) \Xi(t,t') \\
\sin F(t) \delta(t-t') & \cos F(t)\delta(t-t') & \Psi^1(t)h(t') &
\Psi^1(t) \Xi(t,t') \\
0 & 0 & \delta(t-t') - h(t') & - \Xi(t,t') \\
0 & 0 & -h(t') & - \Xi(t,t')
\ea \right)
\eeq
where
\beq
\Xi(t,t') = \Theta(t-t') - \int_{t'}^\infty d\tau h(\tau).
\eeq
Furthermore
\beq
e^i{}_\mu(t,t') = \left( \ba{cccc}
\cos F(t)\delta(t-t') &  \sin F(t)\delta(t-t') & 0 & - \vp^2(t)\delta(t-t')  \\
- \sin F(t)\delta(t-t') &   \cos F(t)\delta(t-t') & 0 & \vp^1(t)\delta(t-t') \\
0            &           0      & \delta(t-t') - h(t) h(t')  & -\delta(t-t') \\
0            &           0      &   0    &  -\partial_t\delta(t-t')    
\ea \right) 
\eeq
as well as 
\beqa
E^\mu{}_i(t,t') & =& \left( \ba{cccc}
\delta(t-t') & 0 & -\Psi^2(t) h(t') & 
- \Psi^2(t) \Xi(t,t') \\
0 & \delta(t-t') & \Psi^1(t)h(t') &
\Psi^1(t) \Xi(t,t') \\
0 & 0 & \delta(t-t') - h(t') & - \Xi(t,t') \\
0 & 0 & -h(t') & - \Xi(t,t')
\ea \right) \no \\
& \cdot& \left( \ba{cccc}
\cos F(t) & -\sin F(t) & 0 & 0 \\
\sin F(t) & \cos F(t) & 0 & 0 \\
0 & 0 & 1 & 0 \\
0 & 0 & 0 & 1 
\ea \right).
\eeqa

\newpage
\begin{figure}
\epsfile{file=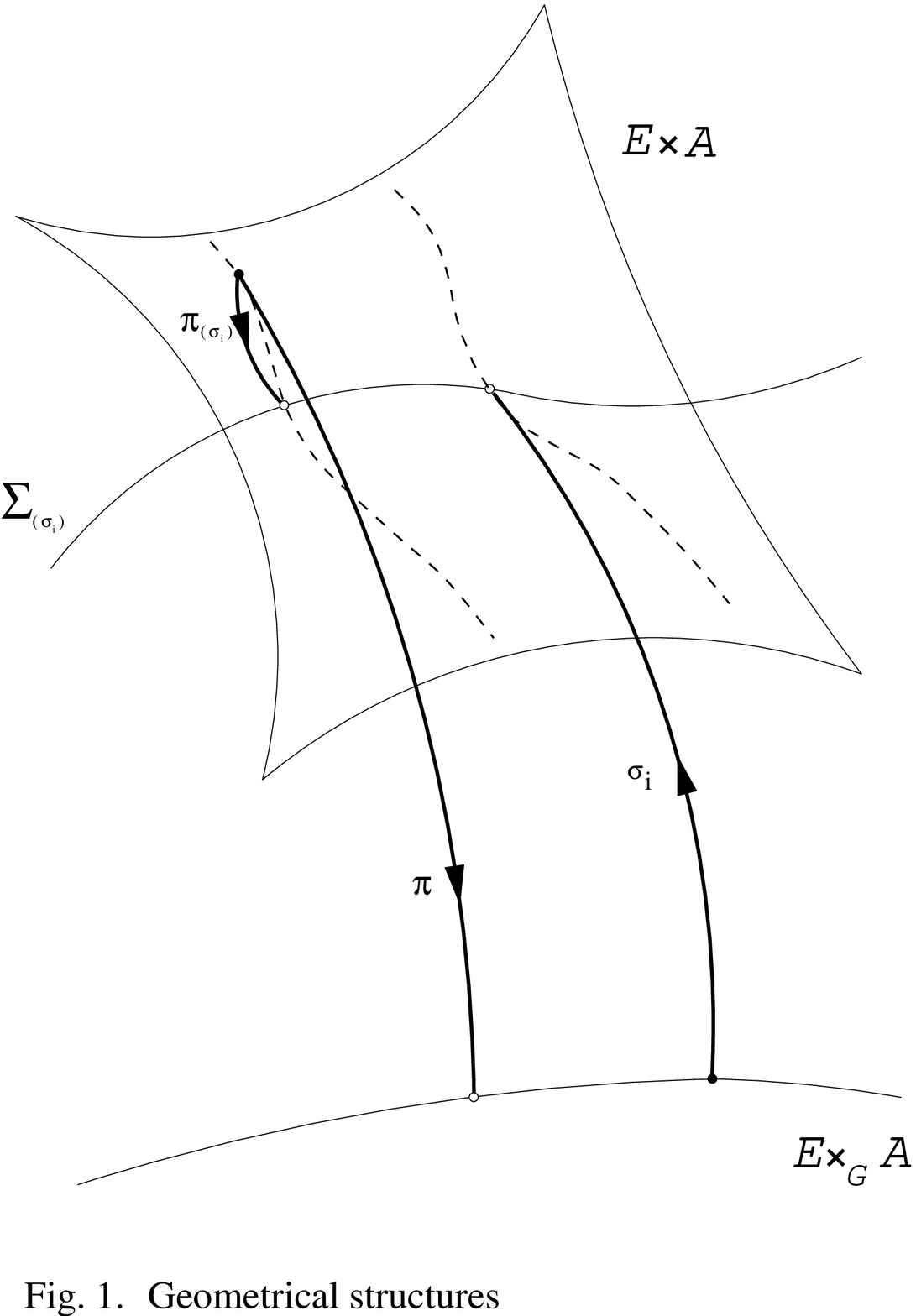}
\end{figure}
\end{document}